\documentclass[twocolumn]{aastex631}

\newcommand{\kms}{\textrm{km s}^{-1}}
\newcommand{\hi}{\text{H\,\sc{i}}}
\newcommand{\Msol}{\textrm{M}_{\odot}}

\newcommand{\VPE}{\ensuremath{V_{PE}}}
\newcommand{\rPE}{\ensuremath{r_{PE}}}
\newcommand{\aPE}{\ensuremath{\alpha_{PE}}}
\newcommand{\Ropt}{\ensuremath{R_{opt}}}
\newcommand{\MI}{\ensuremath{M_I}}
\newcommand{\VHI}{\ensuremath{V_{HI}}}
\newcommand{\RHI}{\ensuremath{R_{HI}}}

\newcommand{\xHI}{\ensuremath{x_{HI}}}
\newcommand{\Vrot}{\ensuremath{V_{rot}}}

\newcommand{\barolo}{\textsc{3DBAROLO}}
\newcommand{\fat}{\textsc{FAT}}
\newcommand{\tirific}{\textsc{TiRiFiC}}
\newcommand{\wkapp}{\textsc{WKAPP}}
\newcommand{\sofia}{\textsc{SoFiA-2}}
\newcommand{\mcgsuite}{\textsc{MCGSuite}}
\newcommand{\ellmaj}{\texttt{ell\_maj}}

\newcommand{\queens}{Department of Physics, Engineering Physics, and Astronomy,Queen's University,\\ Kingston ON K7L~3N6, Canada}
\newcommand{\ICRAR}{International Centre for Radio Astronomy Research (ICRAR), The University of Western Australia,\\ 35 Stirling Highway, Crawley WA 6009, Australia}
\newcommand{\ASTRO}{ARC Centre of Excellence for All Sky Astrophysics in 3~Dimensions (ASTRO~3D),\\ Australia}
\newcommand{\CSIRO}{CSIRO Space and Astronomy, PO Box 1130, Bentley WA 6102, Australia}
\newcommand{\ICRARCURTIN}{International Centre for Radio Astronomy Research (ICRAR) - Curtin University, \\Bentley, WA 6102}

\begin{document}

\title{WALLABY Pilot Survey: Characterizing Low Rotation Kinematically Modelled Galaxies}

\correspondingauthor{N. Deg}
\email{nathan.deg@queensu.ca}

\author[0000-0003-3523-7633]{N. Deg}
\affiliation{\queens}

\author{K. Spekkens}
\affiliation{\queens}

\author[0000-0002-3929-9316]{N. Arora}
\affiliation{\queens}
\affiliation{Arthur B. McDonald Canadian Astroparticle Physics Research Institute, Queen's University, Kingston, ON K7L 3N6, Canada}

\author[0009-0003-3774-3430]{R. Dudley}
\affiliation{David A. Dunlap Department of Astronomy and Astrophysics, University of Toronto, Toronto, ON, M5S 3H4, Canada}
\affiliation{\queens}

\author{H. White}
\affiliation{\queens}

\author[0000-0002-2620-6483]{A. Helias}
\affiliation{Department of Physics \& Astronomy, University of Western Ontario, 1151 Richmond St, London, N6A 3K7, Canada}
\affiliation{\queens}

\author[0000-0001-5310-1022]{J. English}
\affiliation{Department of Physics and Astronomy, University of Manitoba, Winnipeg, Manitoba, R3T 2N2, Canada}

\author{T. O'Beirne}
\affiliation{Centre for Astrophysics and Supercomputing, Swinburne University of Technology, Hawthorn, Victoria 3122, Australia}
\affiliation{\CSIRO}
\affiliation{\ASTRO}

\author[0000-0003-3636-4474]{V. Kilborn}
\affiliation{Centre for Astrophysics and Supercomputing, Swinburne University of Technology, Hawthorn, Victoria 3122, Australia}

\author{G. Ferrand}
\affiliation{Department of Physics and Astronomy, University of Manitoba, Winnipeg, Manitoba, R3T 2N2, Canada}
\affiliation{Interdisciplinary Theoretical and Mathematical Sciences Program (iTHEMS), RIKEN, Wakō-shi, Saitama 351-0198, Japan}

\author[0009-0007-8148-9387 ]{M. L. A. Richardson}
\affiliation{\queens}
\affiliation{Arthur B. McDonald Canadian Astroparticle Physics Research Institute, Queen's University, Kingston, ON K7L 3N6, Canada}

\author[0000-0002-7625-562X]{B. Catinella}
\affiliation{\ICRAR}

\author{L. Cortese}
\affiliation{\ICRAR}

\author[0000-0002-9214-8613]{H. D\'{e}nes}
\affiliation{School of Physical Sciences and Nanotechnology, Yachay Tech University, Hacienda San José S/N, 100119, Urcuquí, Ecuador }

\author{A. Elagali}
\affiliation{School of Biological Sciences, University of Western Australia}

\author{B.~-Q. For}
\affiliation{\ICRAR}

\author{K. Lee-Waddell}
\affiliation{Australian SKA Regional Centre (AusSRC) - The University of Western Australia, 35 Stirling Highway, Crawley WA 6009, Australia}
\affiliation{\CSIRO}
\affiliation{\ICRARCURTIN}

\author{J. Rhee}
\affiliation{\ICRAR}

\author[0000-0003-2015-777X]{L. Shao}
\affiliation{National Astronomical Observatories, Chinese Academy of Sciences, 20A Datun Road, Chaoyang District, Beijing, 100012, China}

\author{A. X. Shen}
\affiliation{\CSIRO}

\author{L. Staveley-Smith}
\affiliation{\ICRAR}
\affiliation{ARC Centre of Excellence for All Sky Astrophysics in 3 Dimensions (ASTRO 3D), Australia}

\author{T. Westmeier}
\affiliation{\ICRAR}

\author{O. I. Wong}
\affiliation{\CSIRO}
\affiliation{\ICRAR}



\begin{abstract}

Many of the tensions in cosmological models of the Universe lie in the low mass, low velocity regime.  Probing this regime requires a statistically significant sample of galaxies with well measured kinematics and robustly measured uncertainties.  WALLABY, as a wide area, untargetted \hi\ survey is well positioned to construct this sample.  As a first step towards this goal we develop a framework for testing kinematic modelling codes in the low resolution, low $S/N$, low rotation velocity regime. We find that the WALLABY Kinematic Analysis Proto-Pipeline (WKAPP) is remarkably successful at modelling these galaxies when compared to other algorithms, but, even in idealized tests, there are a significant fraction of false positives found below inclinations of $\approx40^{\circ}$.  We further examine the 11 detections with rotation velocities below $50~\kms$ in the WALLABY pilot data releases.  We find that those galaxies with inclinations above $40^{\circ}$ lie within $1-2~\sigma$ of structural scaling relations that require reliable rotation velocity measurements, such as the baryonic Tully Fisher relation.  Moreover, the subset that have consistent kinematic and photometric inclinations tend to lie nearer to the relations than those that have inconsistent inclination measures. This work both demonstrates the challenges faced in low-velocity kinematic modelling, and provides a framework for testing modelling codes as well as constructing a large sample of well measured low rotation models from untargetted surveys.

\end{abstract}



\section{Introduction} \label{sec:intro}

There are many tensions between theory and observations of galaxies in the low mass/low velocity regime \citep{Sales2022}.  These range from individual probes of the dark matter (DM) content in individual galaxies, to potential deviations from structural scaling relations like the baryonic Tully Fisher Relation (bTFR), to cosmological statistical measures, like the velocity function, and more.  On an individual basis, there are claims that some gas-rich ultra diffuse galaxies (UDGs) may have little to no DM \citep{ManceraPina2019,ManceraPina2024}.  There are also claims that the larger class of gas-rich UDGs may not lie on the baryonic Tully Fisher Relations (bTFR; \citealt{Mcgaugh2000, Verheijen2001, Hall2012, Lelli2016b,Papastergis2016, Ponomareva2017,Lelli2019,Geha2016,McQuinn2022,Arora2023}) due to their being DM deficient \citep{ManceraPina2020,Hu2023,Du2024}.  It is worth noting that these claims have been disputed in the literature \citep{Brook2021,Sellwood2022,Lelli2024}, and further modelling of some of these galaxies have revealed significantly more DM than initially reported \citep{ManceraPina2022}.  There is further evidence that there may be a downturn in the bTFR at low velocities \citep{McQuinn2022}, but the sample sizes are still relatively sparse.  Beyond these particular examples, there are a number of other tensions in the dwarf galaxy regime (for a full review see \citealt{Sales2022}).  Moving to the larger scale, the velocity function, that is the number of galaxies as a function of rotation velocity, shows a significant discrepancy between observations (using a variety of different techniques to infer the rotation) and theory, with theory predicting far more low rotation galaxies than inferred from observations \citep{Zwaan2010,Papastergis2011,Papastergis15}.

The velocity function is a core prediction and test of cosmology. It is the counterpart to the well-studied luminosity function, which measures the number of galaxies as a function of luminosity.  The luminosity function acts as a cosmological probe of baryonic matter while the velocity function probes the gravitational mass. There have been several observational probes of the velocity function using different tracers.  One approach has been to use the luminosity function coupled with the luminosity-velocity relations like the Tully-Fisher relation \citep{Tully1977} to measure the velocity function (for examples see \citealt{Gonzalez2000,Desai2004}).  It is also possible to utilize the velocity dispersion of early type galaxies \citep{Sheth2003}.  Another approach is to use neutral Hydrogen (\hi) profile linewidths (for example \citealt{Zwaan2010}).  A more recent approach is to use Integral Field Unit (IFU) observations (see for example \citealt{Ristea2024}).  There are also a number of works that combine different tracers together to construct their velocity functions \citep{Klypin2015,Bekeraite2016}. In particular, \citet{Catinella_2023} combined \hi\ and IFU observations to study the Tully Fisher Relation (TFR) and the bTFR. Interestingly, they found a population of low-mass outliers using the IFU observations that are caused by disturbances in the systems.  When examined in \hi, these systems are not outliers and lie along the same relations as the undisturbed galaxies.  A key result of their study is that using IFU data for studying the rotation rates of low mass galaxies may prove to be more challenging than using \hi\ observations. 

For almost as long as the velocity function has been measured observationally, it has also been derived from theory and from large cosmological simulations \citep{Zavala2009,Obreschkow2009,Obreschkow2013,Klypin2015, Dutton2019}.  Comparisons of the observed and predicted velocity functions show a strong degree of tension at the low mass, low velocity end \citep{Zwaan2010,Papastergis2011,Papastergis15}.  There have been various explanations for this difference.  \citet{Abramson2014} posited that one of the drivers is the galaxy type used to construct the velocity function.  \citet{Brooks2017} noted that the \hi\ linewidth may underestimate the true rotational speed and that a lack of sensitivity to low mass dwarfs may lead to an underestimation in observed low velocity galaxies. \citet{Sardone2024} explored the issue of \hi\ profile widths in greater detail and found that the specific definition of peak width is critical for low mass dwarfs.  A core challenge in understanding the velocity function tension is the difficulty comparing simulations to observations.  \hi\ profile widths are often used for observational studies, but the relationship between \hi\ profile widths and the intrinsic properties of the \hi\ disk and dark matter halo (the intrinsic quantities of simulations) is not particularly well defined.

Alleviating the tensions at both the individual galaxy level and at the statistical level requires a statistically significant sample of reliable kinematic models of rotating \hi\ disks.  While many of the explorations of DM-deficient UDGs utilize sophisticated kinematic models, these tend to be poorly resolved with low signal-noise ratios ($S/N)$, and most algorithms have not been extensively tested in this regime.  On the statistical level, there are a number of surveys that have generated relatively large sets of kinematic models. For example, the Spitzer Photometry and Accurate Rotation Curves survey (SPARC; \citealt{Lelli2016b}) compiled observations from a variety of surveys and constructed kinematic models for $\approx200$ galaxies.  The Local Volume \hi\ Survey (LVHIS; \citealt{Koribalski2018}) observed and modelled 82 nearby gas-rich galaxies that were originally detected in HIPASS \citep{Meyer2004}. Many of these models were constructed in \citealt{Kamphuis2015}. \citet{Ponomareva2021} modelled 67 galaxies from the MIGHTEE-HI early science observations \citep{Maddox2021}. Another particularly relevant survey is the Faint Irregular Galaxies GMRT Survey (FIGGS; \citealt{Begum2008a}). These are predominantly low rotation galaxies, and \citet{Begum2008a,Begum2008b} were able to exploit the large spectral resolution of the sample kinematically model a subset of these, reaching rotation velocities down to $\sim20~\kms$.  This list is not exhaustive, but it does contain some of the largest sets of kinematic models to date.  The low sample size of these surveys is insufficient to robustly probe the velocity function, particularly in the low velocity regime.

Kinematic modelling studies of low rotation galaxies using IFU data are more common, but they have shown there are many challenges in this regime.  Difficulties in measuring the inclination as well as correcting for asymmetric drift can affect models \citep{Rhee2004,Valenzuela2007,Oman2016,Pineda2017}.  Moreover non-circular motions \citep{Oman2019a} and non-equilibrium motions due to mergers and gas flow can lead to systematic biases and errors \citep{Downing2023}.  These studies tend to focus on the inner regions of galaxies, but, as shown clearly in \citet{Collins2022}, these effects can affect the outer rotations and bTFR estimates for dwarf galaxies.  On the other hand, (and as noted above) \citet{Catinella_2023} found that IFU observations are more sensitive to disturbances than \hi\ observations of the same systems.  All of this highlights that kinematic modelling in the low velocity regime is challenging and great care must be taken when building samples of low rotation systems.

The Widefield ASKAP L-band Legacy All-sky Blind surveY (WALLABY;  \citealt{Koribalski2020}) will dramatically increase the sample size of kinematically modelled galaxies.  WALLABY is an ongoing untargetted survey with the Australian Square Kilometer Array Pathfinder (ASKAP; \citealt{Hotan2021}) telescope that will cover most of the southern sky, detect the \hi\ content of $>2\times10^{5}$ galaxies, and kinematically model $>10^{4}$ galaxies, an order of magnitude increase over the current literature. The ability to resolve and kinematically model such a large sample of galaxies is a significant step forwards in studies of the resolved \hi\ velocity function, the bTFR, and more.  Rotation curves and kinematic models provide more reliable tracers of rotation than spectral widths and position velocity diagrams.  They can provide measures of the flat portion of a rotation curve and can make measurements at specific isocontours to calculate structural parameters.  WALLABY has already detected the \hi\ content of $\sim2300$ galaxies and provided kinematic models for 236 of these detections using a semi-automated approach \citep{Westmeier2022,Deg2022,Murugeshan2024}.  Eleven of these models have maximal rotation velocities below $50~\kms$.   While these do not constitute a large enough sample for a statistical study, a detailed examination of these galaxies provides a stepping stone towards the measurement of the low mass end of the resolved velocity function. 

To that end, this paper examines modelling in the low resolution,  low $S/N$, low velocity regime in general and those eleven detections in detail. In Sec. \ref{sec:Data} we describe WALLABY and these eleven detections.  Sec. \ref{Sec:LowVelModellingTests} describes the WALLABY kinematic modelling approach and confronts it with the challenges of low-velocity kinematic modelling.  Sec. \ref{sec:Characterization} characterizes the eleven low velocity galaxies using additional optical imaging from the Dark Energy Camera Legacy Survey (DECaLS, \citealt{DESI,Dey2019}) and places these in the context of the larger WALLABY population.  Finally Sec. \ref{Sec:Conclusions} presents our discussion and conclusion.

\section{Data}\label{sec:Data}
\subsection{WALLABY}

WALLABY has an angular and spectral resolution of $30 \arcsec$ and $18.5~\rm{kHz}$ respectively at a noise level of $1.6~\textrm{mJy beam}^{-1}$.  The spectral resolution corresponds to a velocity resolution of $\sim4~\kms$ in the local Universe.   It uses the  \hi\ Source Finding Application (\sofia; \citealt{Serra2015,Westmeier2021}) to detect and parameterize sources in the large datacubes generated for each field observed.
The pilot data releases \citep{Westmeier2022,Deg2022,Murugeshan2024} cover only 1\% of the the target area of WALLABY, their already large data volume has fueled a significant amount of science \citep{Reynolds2021,Reynolds2022,Kim2023,Courtois2023,Glowacki2023,Lin2023,Deg2023b,OBeirne2024,Deg2024,Lee2025}.  

The WALLABY kinematic models are generated via the WALLABY Kinematic Analysis Proto-Pipeline (\wkapp; \citealt{Deg2022}), which is described in more detail in Sec. \ref{ssec:WKAPP}.  The core goal of \wkapp\ is to generate kinematic models for as many WALLABY detections as possible using the \hi\ detections only in order to build a statistically significant sample of robustly modelled galaxies.  As such, all galaxies are modelled in exactly the same fashion rather than tailoring the modelling approach for specific galaxies.  \wkapp\ was tested extensively in \citet{Deg2022}, but that testing focused on its reliability for the majority of the WALLABY galaxies and the specific regime of low resolution, low $S/N$, low rotation detections was not explored.  As such, it is important to both examine the subset of low rotation galaxies found in WALLABY and test more modelling algorithms in the low rotation regime more generally.

\begin{figure}
\centering
    \includegraphics[width=0.9\columnwidth]{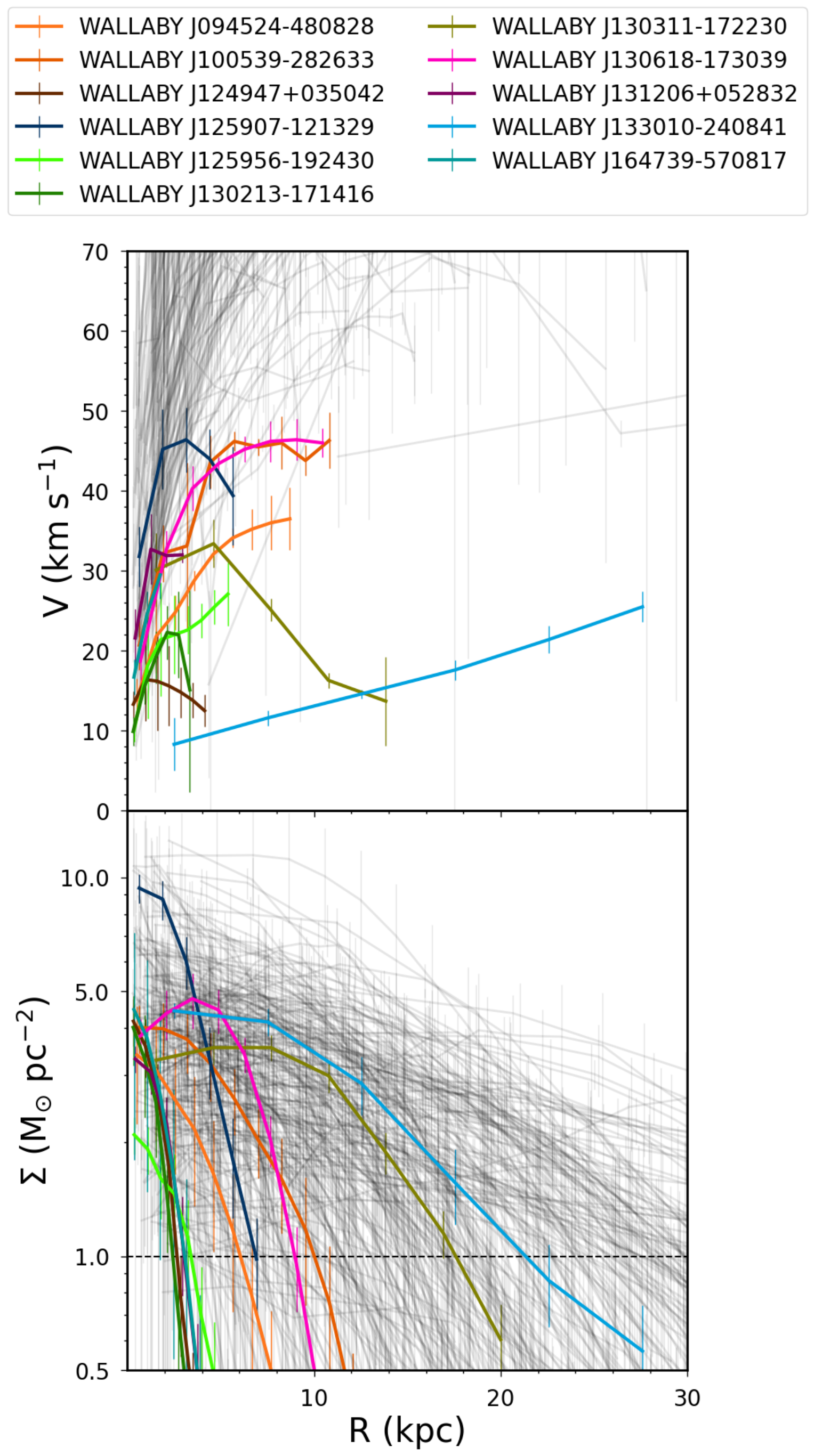} 
    \caption{The rotation curves (top panel) and surface density profiles (bottom panel) for the 11 low-velocity kinematic models.  In each panel, the gray lines show the entire suite of WALLABY PDR kinematic models.}
  \label{Fig:LowVel_RCs}
\end{figure}

To select a low velocity sample from the released WALLABY kinematic models, we adopted a limit of $V_{\rm{max}}\le50~\kms$, where $V_{\rm{max}}$ is the maximum rotation velocity of the model.  This $50~\kms$ limit is located in the range where models \citep{Chauhan2019} and observations \citep{Papastergis2011} for the \hi\ velocity function show significant tension. Figure \ref{Fig:LowVel_RCs} shows the rotation curves and surface density profiles for these 11 galaxies compared to the full suite of WALLABY PDR kinematic models. These galaxies typically have smaller radial extents than the larger WALLABY population (with the exception of WALLABY J133010-240841 and WALLABY J130311-172230). While their rotation curves are low by definition, their central surface densities and the rate at which the profiles decrease with radius are comparable to the larger WALLABY population. Some galaxies (WALLABY J130311-172230 in particular) seem to have falling rotation curves, but these may due to noise or issues in the modelling (see Sec. \ref{sec:Characterization}).  Falling rotation curves are indeed interesting, but, given the uncertainties, a full investigation of these features is beyond the scope of this work.  It is also worth noting that WALLABY J130311-172230 and WALLABY J130618-173039 show evidence of holes towards the center of their surface density profiles.  These holes appear to be real (see below), and not due to issues in the centering.  Nonetheless, while there is a decrease in the gas surface density in that inner region, there is still sufficient gas to calculate the rotation curves.  That being said, the inner rotation curves and surface density profiles will be strongly affected by beam smearing. 

Table \ref{tab:LowVel_Basics} lists these low velocity detections.  The table lists the names, the WALLABY field of the observation, cross-matched IDs and sky coordinates.  In the fields column, NGC 5044 TR2 and NGC 5044 TR3 correspond to the second and third NGC 5044 source finding releases, and Hydra TR2 corresponds to the second Hydra source finding release.  For each detection, we have searched for a corresponding optical detection in DECaLS.  The optical bands column lists each of the $g, r, z$ bands with available imaging from DECaLS. The classification column lists the morphology of the galaxy (see Sec. \ref{sec:Characterization} for more details).  WALLABY J094524-480828 and WALLABY J164739-570817 are not found in DECaLS, but there is shallow archival data for them in the Digital Sky Survey (DSS, \citealt{Lasker1994,McLean2000}).  While these are subject to significant extinction, there are HyperLEDA classifications available \citep{Makarov2014}, which we include in Table \ref{tab:LowVel_Basics}.

Table \ref{tab:DerivedProperties} further lists properties of the low velocity galaxies derived from their observations in \citet{Westmeier2022,Murugeshan2024}, and the analysis of their structural parameters in \citet{Deg2024}. In \citet{Deg2024}, the optical inclinations are derived using the \textsc{AutoProf} package \citep{Stone2021}, and the stellar masses are derived using an approach based on \citet{Arora2021} (see \citet{Deg2024} for a detailed explanation of their derivation).  However,  WALLABY J130213-171416 and WALLABY J130311-172230 do not have $r$ band images in DECaLS, which were required in \citet{Deg2024}.  In this work, we have relaxed our requirements and measured the optical inclination and stellar mass for these two galaxies using only the  $g$ and $z$ bands (which leads to increased uncertainties).  For WALLABY J094524-480828 and WALLABY J164739-570817, we have adopted the HyperLEDA inclinations, but were unable to make measurements of their stellar mass.

\begin{deluxetable*}{ c c c c c c c }
\tablewidth{1.0\textwidth} 
\tablehead{
\colhead{WALLABY Name}       & \colhead{WALLABY Field}       &  \colhead{Cross match ID}     & \colhead{RA}       & \colhead{DEC}   & DECaLS Bands & Classification 
}
\startdata
\hline
 &  &  & $^{\circ}$ & $^{\circ}$  \\
\hline 
WALLABY J130213-171416 & NGC 5044 TR3 & UGCA 319 & $195.557\pm0.005$&$-17.2365\pm0.0009$ & g, z & D\\
WALLABY J125907-121329 & NGC 5044 TR3 & UGCA 312& $194.7790\pm0.0012$ &$-12.2290\pm0.0005$ & g, r, z & D\\
WALLABY J125956-192430 & NGC 5044 TR3 & LEDA 44681 & $194.9813\pm0.0001$&$-19.4100\pm0.0019$ & g, r, z & D\\
WALLABY J131206+052832 & NGC 4808 & UGC 8276 & $198.0274\pm0.0002$ &$5.4753\pm0.0008$0  & g, r, z& Irr\\
WALLABY J130618-173039 & NGC 5044 TR3 & MCG-03-33-032 & $196.5756\pm0.0001$& $-17.5105\pm0.0003$ & g, r, z& Irr\\
WALLABY J100539-282633 &  Hydra TR2 & ESO 435-39 &$151.4132\pm0.0007$& $-28.4450\pm0.0008$ & g, r, z& Pair\\
WALLABY J124947+035042 &  NGC 4808 & UGC 7983 & $192.4491\pm0.0006$& $3.8422\pm0.0012$ & g, r, z & Sp\\
WALLABY J130311-172230 & NGC 5044 TR2 & MCG-03-33-029& $195.7988\pm0.0001$& $-17.3756\pm0.0008$ & g, z& Sp\\
WALLABY J133010-240841 & NGC 5044 TR2 & ESO 509-33 & $202.5450\pm0.0002$&$-24.1435\pm0.0012$ & g, r, z& Sp\\
WALLABY J094524-480828 & Vela & PGC 2791980 & $146.3518\pm0.0004$ & $-48.1412\pm0.002$ & - & Sp\\
WALLABY J164739-570817 & Norma & PGC 393557 & $251.9156\pm0.0002$& $-57.1379\pm0.0006$ & - & Irr\\
\hline
\enddata
\caption{Properties of kinematically modelled WALLABY PDR galaxies with maximum rotations $<50~\kms$.  RA and DEC are from \citet{Deg2022} and \citet{Murugeshan2024}.  In this table, D==dwarf,  Irr== Irregular, and Sp == Spiral.}
    \label{tab:LowVel_Basics}
\end{deluxetable*}

\begin{figure*}
\centering
    \includegraphics[width=0.9\textwidth]{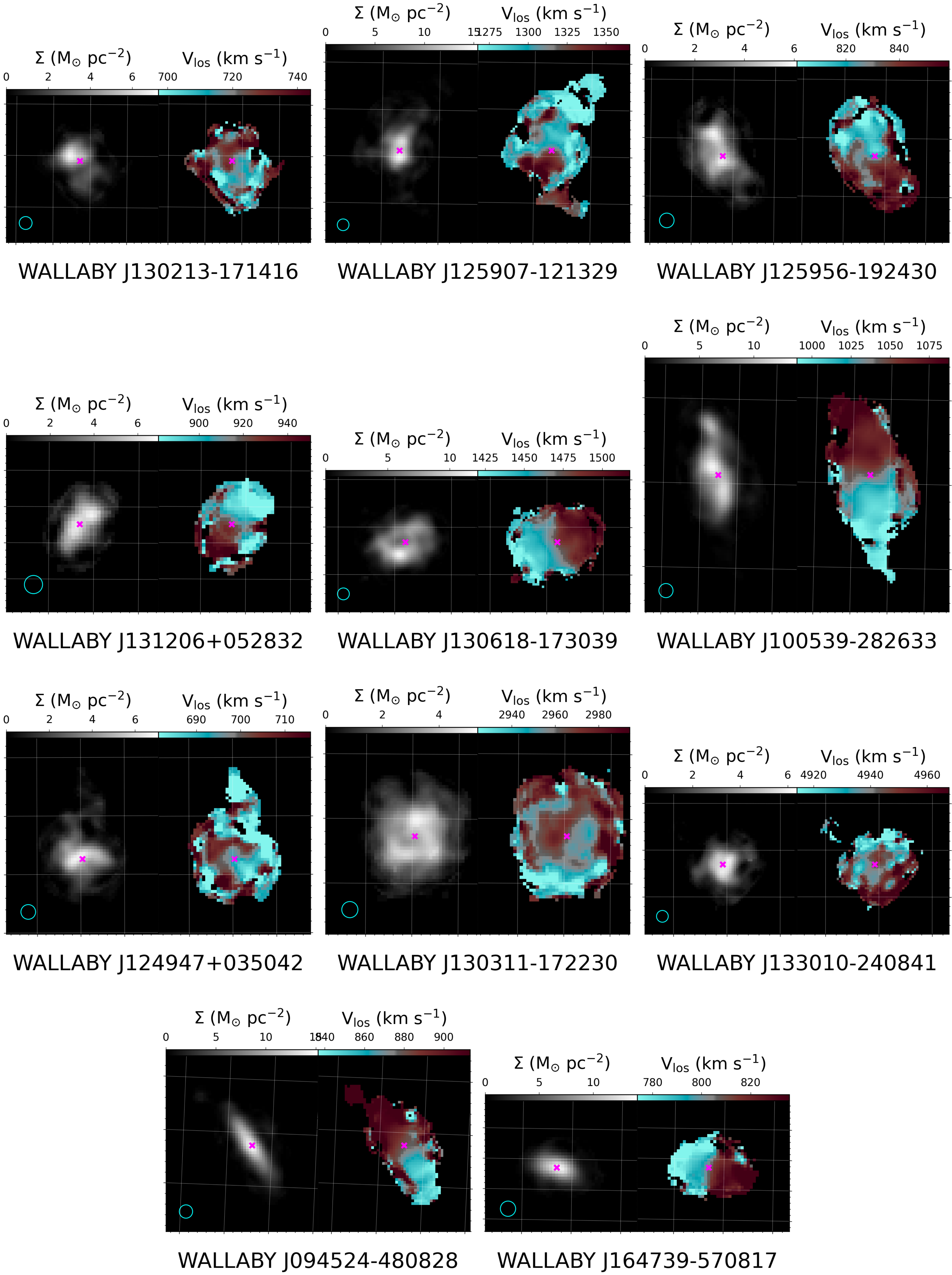} 
    \caption{The moment 0 \& 1 maps for the 11 low velocity WALLABY detections.  The moment 1 map uses a custom color scheme from \textsc{CosmosCanvas} \citep{English2024} designed so that the approaching, blue shifted gas appears to come out from the page, while the red shifted gas goes into the page.  The cyan circle in the moment 0 maps show the size of the beam, and the magenta x shows the location of the kinematic center listed in Table \ref{tab:LowVel_Basics}.  The light gray grid illustrates lines of RA-DEC relative to the map coordinates.  The moment 0 maps use a linear scale which causes them to appear to be smaller in size than the moment 1 maps, but the two do indeed cover the same parts of the sky.  The panels are organized by optical type, with the top row containing the dwarf-like galaxies, the 2nd row containing the irregular detections, the third row containing the face-on spirals, and the bottom row containing those detections not found in DECaLS}.
  \label{Fig:MomentMaps}
\end{figure*}

Figure \ref{Fig:MomentMaps} shows moment 0 and moment 1 maps of the low velocity sample.  A brief examination of these systems shows that these detections have a significant degree of structure.  The moment 1 maps all show signs of rotation, but many of them, particularly WALLABY J130311-172230 (3rd row, middle column), also show significant evidence of non-circular motions as well.  These non-circular motions are generally seen in the low column density outskirts, where the noise and potential errors in the underlying \sofia\ masks used to make the maps may combine to emphasize departures from symmetry. Since the underlying codes used in \wkapp\ (see Sec. \ref{ssec:WKAPP}) fit models from the inside out, these non-circular motions will only affect the outermost regions of the models. Most of the galaxies show regular gas distributions in the central regions.  But, there are decreases in the moment 0 maps for WALLABY J130311-172230 and WALLABY J130618-173039, which are reflected in the surface density profiles shown in Figure \ref{Fig:LowVel_RCs}.  These holes occur near, but not directly at, the center of the gas distribution.  Other than noting these holes, it is difficult to draw any further conclusions due to the low resolution, low $S/N$, and effect of beam smearing.  A full investigation of their origin will require higher sensitivity and resolution observations.  Nonetheless, these will not affect the rotation curve derivation as that relies on the full 3D distribution.

\begin{deluxetable*}{ c c c c c c c c c}
\tablewidth{1.0\textwidth} 
\tablehead{
\colhead{WALLABY Name}        & \colhead{V$_{\rm{max}}$}      & \colhead{V$_{\rm{sys}}$}       & \colhead{Dist}    &  \colhead{Inc$_{\rm{kin}}$}  &  \colhead{Inc$_{\rm{opt}}$}  & \colhead{log$_{10}$(M$_{\hi}$/M$_{\odot}$)}    & \colhead{log$_{10}$(M$_{*}$/M$_{\odot}$)}
}
\startdata
\hline
 & $\kms$ & $\kms$ & Mpc & $^{\circ}$ & $^{\circ}$  \\
\hline 
WALLABY J130213-171416 & $22\pm3$ & $720.6\pm3.5$&  8 & $29\pm2$ & $58$ & $7.87\pm0.03$& $7.9 \pm 0.1$\\
WALLABY J125907-121329 & $46\pm4$ & $1316.4\pm0.9$& 17 & $41\pm4$ & $53$ & $8.95\pm0.02$& $8.1\pm0.2$\\
WALLABY J125956-192430 & $27\pm3$ & $830.1\pm1.5$& 10 & $65 \pm 2 $ & $47 $& $8.10\pm0.03$& $7.2\pm0.2$\\
WALLABY J131206+052832 & $33\pm4$ & $916.0\pm0.8$ &12 & $56\pm3$& $65$& $8.05\pm0.04$& $7.6\pm0.1$\\
WALLABY J130618-173039 & $46\pm3$ & $1468.9\pm0.4$&19 &$50\pm1$ & $52$&$9.04\pm0.01$ &$8.4\pm0.1$ \\
WALLABY J100539-282633 & $46\pm 4$ & $1039.0\pm1.1$& 17& $ 68 \pm 3$ & $71$ & $8.99\pm 0.01$ & $8.2\pm 0.2$\\
WALLABY J124947+035042 & $16\pm5$ & $698.6\pm1.8$ &9 & $45 \pm 5$ & $27$& $8.00\pm0.02$ & $7.4\pm0.1$\\
WALLABY J130311-172230 & $33\pm3$ & $2959.3\pm0.5$ & 42 & $26\pm 1$ & $33$ & $9.55\pm0.02$& $8.9 \pm 0.1$\\
WALLABY J133010-240841 & $26\pm2$ & $4940.9\pm1.1$& 69&$36\pm1$ &$28$ & $9.78\pm0.02$& $9.9\pm0.1$\\
WALLABY J094524-480828 & $37\pm4$ & $874.9\pm0.4$& 14& $76 \pm 5$& $\sim 90$ & $8.62\pm0.01$ & - \\
WALLABY J164739-570817 & $30\pm3$ & $804.8\pm0.5$& 10 & $70\pm12$& $\sim 80$ &$8.04\pm0.02$ & - \\
\hline
\enddata
\caption{Derived properties of of the low-velocity sample from WALLABY PDR.  $V_{\rm{max}}$, $V_{\rm{sys}}$, and Inc$_{\rm{kin}}$ are from \citet{Deg2022} and \citet{Murugeshan2024}, while the distance, Inc$_{\rm{opt}}$, \hi\ and stellar masses are from \citet{Deg2024}.  The inclinations for WALLABY J094524-480828 and WALLABY J164739-570817 are from the HyperLEDA catalogue.}
    \label{tab:DerivedProperties}
\end{deluxetable*}

\section{Modelling in the Marginally Resolved, Low Resolution, and Low Velocity Regime} \label{Sec:LowVelModellingTests}

The 236 WALLABY PDR kinematic models released in \citet{Deg2022} and \citet{Murugeshan2024} are generated using the WALLABY Kinematic Analysis Proto-Pipeline (\wkapp).  This algorithm combines fits from both the 3D-Based Analysis of Rotating Objects From Line Observations (\barolo; \citealt{diTeodoro15}) code and the Fully Automated \tirific\ (\fat, where \tirific\ itself stands for Tilted Ring Fitting Code; \citealt{Kamphuis2015, Jorza2007}) code. These are both 3D Tilted Ring (TR) modelling codes that generate their models by fitting 3D data cubelets.  \wkapp\ was optimized for low resolution and low $S/N$ WALLABY detections.  However, neither it, nor \fat\ or \barolo, have been optimized and tested extensively in the low resolution, low $S/N$, \textit{low velocity} regime that describes the eleven galaxies in our sample.  This regime is among the most challenging for kinematic modelling for a variety of reasons.  Firstly, at low rotation velocities, the influence of even low velocity dispersions and non-circular motions on the dynamics is significant.  Secondly, low rotation disk galaxies will also have lower masses, making them more susceptible to disruptions and interactions.  Thirdly, due to their low mass, they are more easily detected in face-on orientations, which itself is challenging for kinematic modelling (see \citet{Deg2022} for a more in depth exploration).  Fourthly, reliable kinematic models require robust inclination measurements, either from the model itself or from optical imaging.  But, when galaxies are marginally resolved, the kinematic inclination can be challenging to measure.  And, while the optical inclination may be more reliable, at low masses, the stochastic nature of star formation can affect the measurement of the inclination.  Given these challenges, it is critical to test the reliability of modelling in the low resolution, low $S/N$, low velocity regime in order to explore the reliability of our 11 low velocity models.  Sec. \ref{ssec:WKAPP} describes the \wkapp\ algorithm and Sec. \ref{ssec:LowVelTesting} presents the results of our testing.

\subsection{WKAPP}\label{ssec:WKAPP}

The core idea of \wkapp\ is to generate a kinematic model by averaging together \barolo\ and \fat\ fits.  As shown in \citet{Deg2022}, this produces significantly more robust models than either code alone.  A full description of \wkapp\ is available in \citet{Deg2022}.  In brief, the steps involved in generating a \wkapp\ model for WALLABY are:
\begin{enumerate}
    \item Convert WALLABY source cubelets from frequency space to velocity space and smooth them to a 12 $\kms$ resolution via \barolo;
    
    \item Select all detections with a \sofia\ estimated size, \ellmaj\ (the approximate diameter of the major axis), greater than 2 beams as well as all detections with an integrated $\textrm{log}_{10}(S/N)>1.25$ (see Equation 1 of \citealt{Deg2022}) for fitting with both the \barolo\ and \fat\ fitting codes;

    \item Visually examine the \barolo\ and \fat\ fits for each detection.  If both fits are successful, visually agree and satisfy $|\sin(i_{\rm{\fat}})-\sin(i_{\rm{\barolo}})|\le 0.2$ generate the averaged model (i.e proceed to step 4);
    
    \item To generate a \wkapp\ model, start by averaging the \fat\ and \barolo\ center points, inclination, position angle, and systemic velocity.  The uncertainty is set to half the difference of the two fits;
    
    \item Correct both the \fat\ and \barolo\ rotation curves to the averaged inclination and then average those two together to obtain a measured rotation curve.  Use the half the difference between the fits as the uncertainty;
    
    \item Calculate the \wkapp\ surface density by ellipse fitting the masked moment 0 map using the kinematic geometry.  Use the rms of the pixels in the ellipse to get the surface density uncertainty.
\end{enumerate}

There are a few key things to note about \wkapp\, \fat\, and \barolo.  Despite both being 3D TR modelling codes, \fat\ and \barolo\ use different source finders, minimizers, and approaches to calculating the likelihood.  In the low $S/N$ space of WALLABY detections the likelihood space is also fairly flat.  Thus, it is unsurprising that, in many cases, \fat\ and \barolo\ yield fairly different models.  This is made even more challenging by the fact that the initial estimates can strongly affect the final model (see for example Figure 2 in \citealt{Deg2022}).  In \wkapp, both \fat\ and \barolo\ are run in similar fashions (using `flat-disk' geometries with constant inclinations and position angles and providing no initial estimates).  \citet{Deg2022} noted that the differences between the \fat\ and \barolo\ runs provided a better representation of the uncertainty than either the built-in error estimates of \barolo\ or \fat.  While the \wkapp\ uncertainties do not truly represent the underlying distribution of available models, they are a best estimate and are more reasonable than either the \barolo\ or \fat\ uncertainties.  For further detail see Secs. 3.2 and 4.6 of \citet{Deg2022}.

\wkapp\ was validated for the majority of WALLABY detections, but its reliability in the low velocity regime has yet to be explored.  This is relevant as there is reason to suspect that the spectral averaging may affect the reliability of models in this regime due to the relatively low volume they occupy in the cubelets.  For example, a galaxy with a maximum rotation velocity of $50~\kms$ will span $\sim10$ channels, while a $30~\kms$ detection will only span $\sim6$ channels for \wkapp.  The remainder of this section will be focused on assessing the reliability of \wkapp\ in this regime.

\subsection{Testing WKAPP}\label{ssec:LowVelTesting}

We generated two distinct suites of models to test \wkapp\ (as well as \barolo\ and \fat) in the low velocity, low resolution, low $S/N$ regime using the MockCubeGenerator Suite (\mcgsuite\; \citealt{Lewis2019}) code.  \mcgsuite\ is designed to generate idealized mock cubelets for testing.  The code is publicly available at \href{https://github.com/CIRADA-Tools/MCGSuite}{github} and a full description is provided in Appendix \ref{app:MCGSuite}.  It has been used in \citet{Deg2022} for some of the initial tests of \wkapp, in \citet{Deg2023a} to study asymmetries, in \citet{Deg2023b} to study polar rings, and in \citet{Gasser2023} to study warps.

The first test suite consists of 100 rotating disks generated using the \texttt{make\_suite\_MCG\_random} \mcgsuite\ wrapper (see Appendix \ref{app:MCGSuite} for a description of this wrapper).  All galaxies in this mock sample have \hi\ diameters, $D_{\hi}$ between $2\le D_{\hi}~\rm{(beams)}\le 8$, $7.5 \le \log_{10}(M_{\hi}/\Msol)\le 8.5$, $5^{\circ}\le \rm{Inc}\le 90^{\circ}$, $6 \le V_{\rm{disp}}/\kms \le 8$, and a noise $\sigma=1.6~\rm{mJy/Beam}$.  The mass range of this sample yields a range of rotation velocities between $\sim 20$ and $\sim 70~\kms$, which is comparable to our low velocity WALLABY sample.  The second test suite consists of 100 `velocity decoupled' (VD for short) models where the rotation curve does not depend on the input mass (see Appendix \ref{app:MCGSuite}.  This suite contains cubelets with the same mass, surface density, and rotation velocity range as the `normal' suite.  The motivation for this second suite is that there is evidence that some low mass systems may not follow the scaling relations used to build the `normal' \mcgsuite\ galaxies \citep{ManceraPina2019,ManceraPina2020}.  The VD suite ultimately explores a wider range of parameter space in terms of the underlying rotation curves and surface density profiles. 

Once generated, each suite is fit using \wkapp.  As a part of step 3 in \wkapp, the individual \fat\ and \barolo\ fits are visually examined in isolation to determine if they are acceptable (and if they agree with each other). These fits may be unacceptable if a) the specific code fails to generate a model altogether, or b) if the fit contains unphysical features.  Both \fat\ and \barolo\ must produce an acceptable fit \textit{and} the two fits must agree with each other in order for \wkapp\ to produce a model (see Sec. \ref{ssec:WKAPP}).  It is after the generation of the \wkapp\ models that all three fits (\wkapp\, \barolo\, and \fat) are compared to the initial \mcgsuite\ models.  In this second comparison, the fits are visually examined to determine if they recover the input model. This comparison is similar to the \wkapp\ visual comparison between \barolo\ and \fat\ that is used to determine whether the two fits agree.  We will note here that this is not a quantitative comparison due to both the large parameter space and the challenges of the \barolo, \fat\, and \wkapp\ uncertainties.  However, this idea has been explored in \citet{Lewis2019}, who attempted a variety of acceptability quantifications and found that none of their attempts fully captured the nuances of acceptability.  Appendix \ref{app:TestingPlot} shows a pair of diagnostic plots to illustrate the difference between fits that agree and those that disagree with the underlying models.

\begin{figure*}
\centering
    \includegraphics[width=0.9\textwidth]{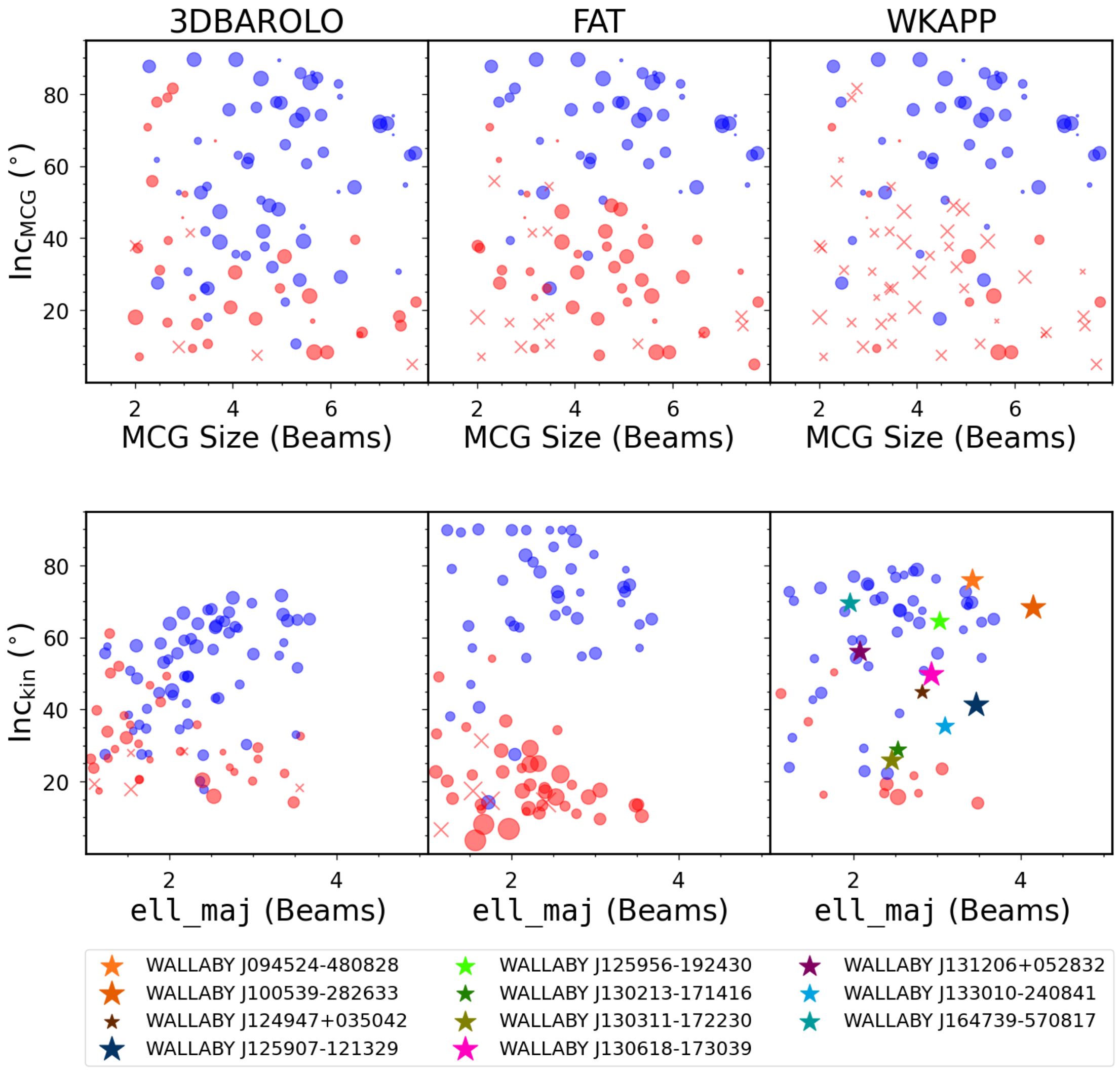} 
    \caption{The success and failures of \barolo\, \fat\, and \wkapp\ for a sample of 100 idealized low velocity galaxies generated by \mcgsuite.  Circles and X's indicate whether the models are accepted based only upon a visual inspection of their outputs.  The colors of the symbols indicate whether (blue) or not (red) the models match the underlying \mcgsuite\ model for a particular mock cubelet.  In other words red circles would contaminate a final catalogue.  The panels in the upper row show the \mcgsuite\ size and inclination, with the size of the symbols mapping to the input \hi\ mass, while the lower row shows \ellmaj (the \sofia\ size estimate) and modelled inclination, with the size of the symbols mapping to the modelled $V_{\rm{max}}$.  The 11 low-velocity WALLABY detections are shown in the bottom right panel.}
  \label{Fig:LowVel_MCGResults}
\end{figure*}

Figure \ref{Fig:LowVel_MCGResults} shows the results for the sample of `normal' low velocity mock cubelets.  Circles and X's indicate whether a particular fit is visually acceptable, while the blue and red colors indicate whether the model agrees with the initial \mcgsuite\ model.  Ideally, a final catalogue generated by a kinematic modelling algorithm would consist of only fits that agree with the underlying model (i.e. blue circles).  However, there are always models that appear to be reasonable based on a visual inspection but do not agree with the underlying truth (the red circles).  These contaminate any final catalogue of kinematic models as there is no \textit{a priori} reason to reject them until they are compared with the underlying truth.

The upper row of Figure \ref{Fig:LowVel_MCGResults} shows the input \mcgsuite\ size and inclination.  Since the suite is the same for all three codes, each realization is at the same point in each of the three upper panels.  It is clear that all the codes are more successful recovering the input \mcgsuite\ models at higher inclinations than lower inclinations.  Below an input inclination of $\sim 40^{\circ}$, \barolo\ is successful $\approx 50\%$ of the time, while \fat\ is rarely successful.  Both codes have a relatively high ratio of red/blue circles (32/62 and 41/44 for \barolo\ and \fat\ respectively).  By contrast, the requirement that both \fat\ and \barolo\ must agree for \wkapp\ to produce a model means that there are very few interlopers compared to the other codes (\wkapp\ has a red/blue circle ratio of 11/45).  The trade-off is that \wkapp\ has fewer accepted models than either \fat\ or \barolo.

The bottom row of Figure \ref{Fig:LowVel_MCGResults} shows the same suite of models using their estimated size from \sofia\ and measured inclination, with the size of the symbols corresponding to the modelled $V_{\rm{max}}$.  These panels show fewer models than the upper panels as \barolo\ and \fat\ do not always return a model (one of the possible causes of the red X's in the upper panels).  For this reason there are no red X's in the lower \wkapp\ panel. Comparing the \barolo\ and \fat\ panels, it is clear that they have different limits on the measured inclinations. \fat\ will set the inclination to $90^{\circ}$ when the minor axis is resolved by a single resolution element.  By contrast, no marginally resolved \barolo\ model reaches an inclination above $80^{\circ}$.  Similarly, \fat\ can return lower inclinations than \barolo\ due to various checks in the two codes.  \wkapp\ averages the \fat\ and \barolo\ inclinations, which explains why the inclination distribution lies between the other two.  It is worth noting that the \ellmaj\ measure is systematically lower than the input MCG size, which itself is $R_{\hi}$.  This is a known property of \sofia\ due to the noise limit (see \cite{Deg2022} for a comparison of \ellmaj\ and $R_{\hi}$).

In the bottom row of Figure \ref{Fig:LowVel_MCGResults}, there are few blue circles below $~\sim 40^{\circ}$ and almost no successes below $25^{\circ}$ regardless of the code used.  For \wkapp\ the ratio of red/blue circles is only marginally better than $50\%$ in the range between $20^{\circ}-40^{\circ}$.  In other words, even in idealized cases, models with maximum velocities below $50-60~\kms$ and inclinations below $40^{\circ}$ should be approached with a great deal of caution.

\begin{figure*}
\centering
    \includegraphics[width=0.9\textwidth]{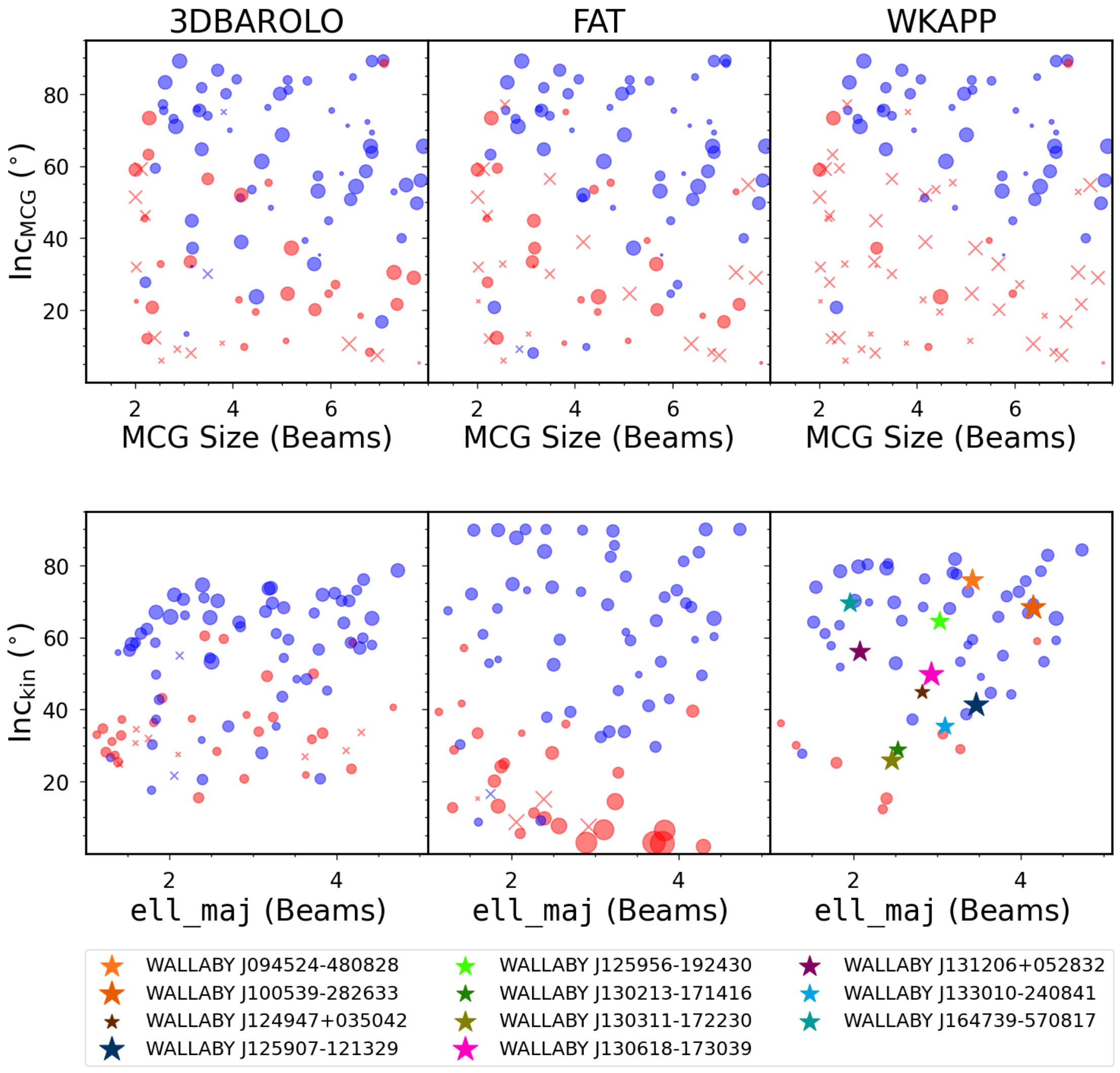} 
    \caption{The success and failures of \barolo\, \fat\, and \wkapp\ for a sample of 100 idealized VD galaxies generated by \mcgsuite.  Panels and symbols are the same as those in Figure \ref{Fig:LowVel_MCGResults}.}
  \label{Fig:UDG_MCGResults}
\end{figure*}

Figure \ref{Fig:UDG_MCGResults} shows the results for the mock VD suite.  As with Figure \ref{Fig:LowVel_MCGResults}, the large fraction of red circles below measured inclinations of $\approx40^{\circ}$ indicates that real observed models in this regime should be considered to be unreliable.  There is one additional thing to note about Figure \ref{Fig:UDG_MCGResults}.  There are a small set of blue X's seen in the \barolo\ and \fat\ panels.  These are fits that were rejected based on the initial visual inspection but actually match in the input \mcgsuite\ model.  This simply highlights the fact that, in a pipeline based approach, some good fits will always end up rejected from the final catalogues.

One possible driver of moderate inclination failures is the spectral resolution.  As noted in \citet{Begum2008a}, a high spectral resolution facilitates the modelling of low rotation galaxies.  By contrast, in \wkapp, the cubelets are first smoothed to a 12 $\kms$ spectral resolution and then fit.  At these low velocities, a galaxy will span only a few channels, which may affect the fitting.  In order to test whether this may drive some of the failures, we re-fit the VD sample using the 4 $\kms$ resolution cubelets. We chose this suite due to the wider range of possible rotation curve and surface density profiles. We also supplemented the 100 original mock galaxies with an additional 40 \mcgsuite\ mocks with sizes ranging from 8-15 beams.  These are intended to determine if a higher spatial resolution may also improve the fitting results at low inclinations.

\begin{figure}
\centering
    \includegraphics[width=0.9\columnwidth]{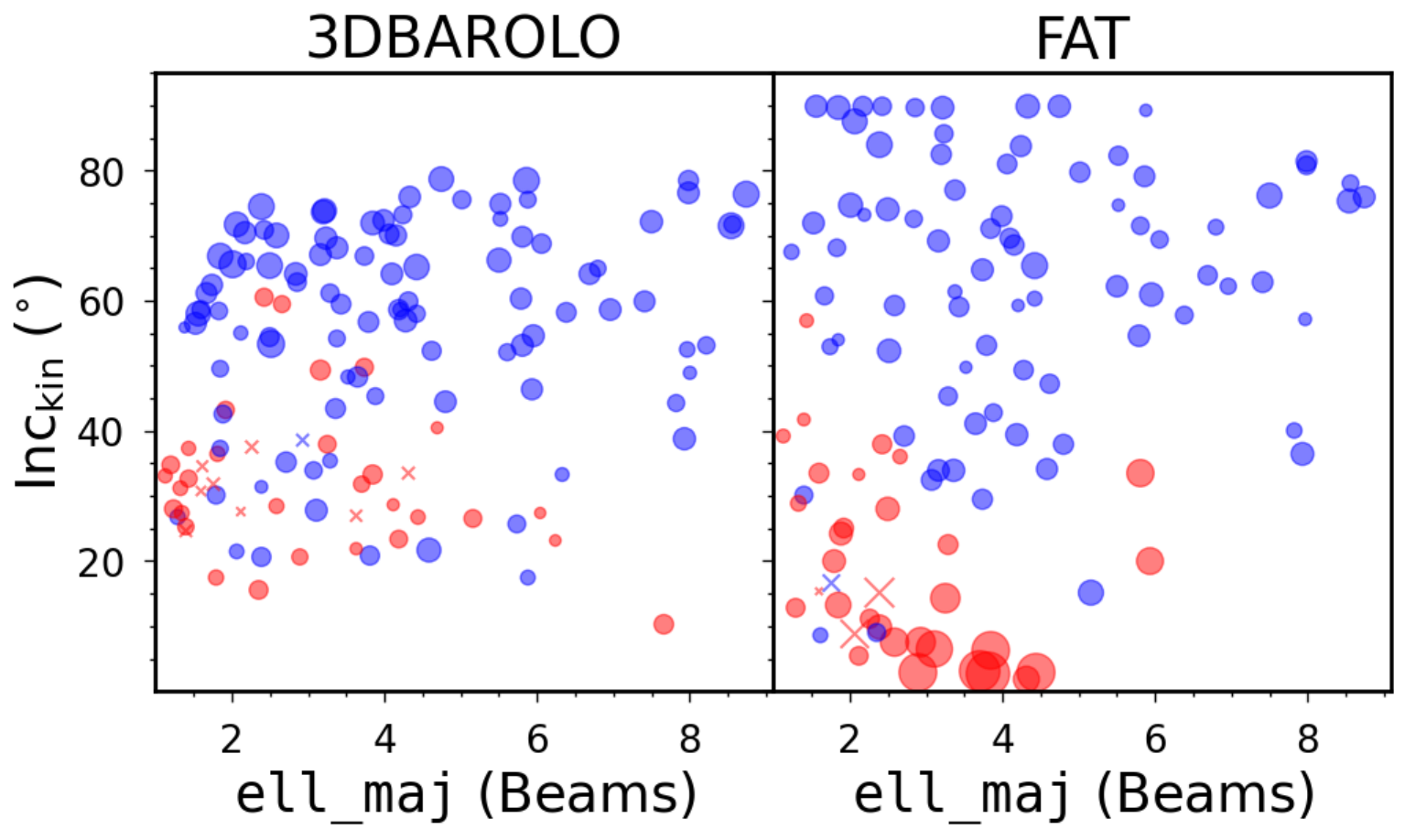} 
    \caption{The results of fitting the extended VD suite of models at $4~\kms$ spectral resolution for \barolo\ and \fat.  Symbols and colors are the same as Figure \ref{Fig:LowVel_MCGResults}.}
  \label{Fig:UDG_MCGResults_HighRes}
\end{figure}

Figure \ref{Fig:UDG_MCGResults_HighRes} shows the results from this extended analysis of higher resolution detections. There is little change in the \fat\ results between the 4 and 12 $\kms$ results.  Nor is there an improvement in the \barolo\ fits.  Regardless of the resolution, both codes struggle at moderate to low inclinations.  We caution against generalizing here, as these struggles are specific to low resolution, low $S/N$, low velocity WALLABY-like galaxies.  At the higher velocities and higher $S/N$ of many WALLABY galaxies, these struggles do not exist \citep{Deg2022}.  But, it is clear that great care must be taken when modelling galaxies with low rotation velocities with moderate inclinations.

\begin{figure*}
\centering
    \includegraphics[width=0.9\textwidth]{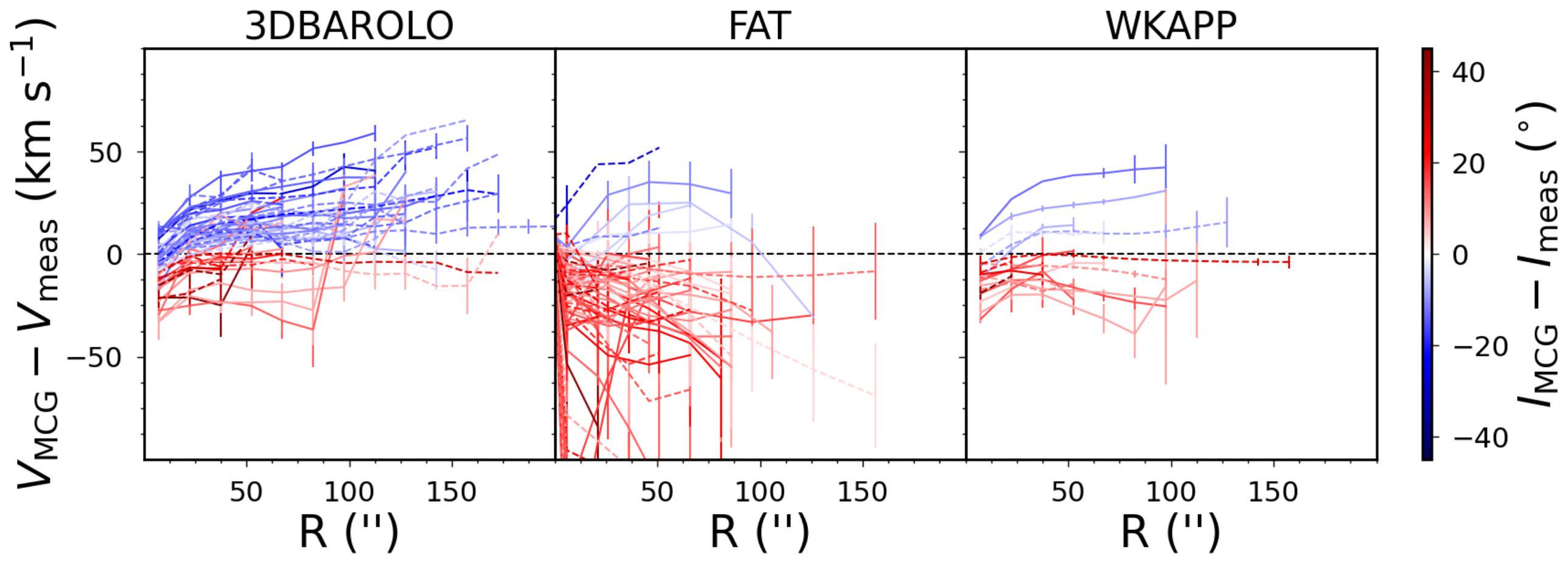} 
    \caption{Biases in the rotation curves from fits to the sample of low mass galaxies that do not agree with the initial \mcgsuite\ model. Solid and dashed lines are for the normal and VD populations respectively.  And the line colors show the difference between the true and measured inclination.}
  \label{Fig:Inc_Rot_Bias}
\end{figure*}

Figure \ref{Fig:Inc_Rot_Bias} shows the differences between the true and measured rotation curves for accepted galaxies that do not agree with the underlying \mcgsuite\ models (i.e. the red circles in Figures \ref{Fig:LowVel_MCGResults} and \ref{Fig:UDG_MCGResults}).  To be clear, each residual shown in Figure \ref{Fig:LowVel_MCGResults} would be a contaminant in the final catalogue of accepted models as this figure does not show the residuals for accepted models that match the underlying \mcgsuite\ model.  In this plot, there does not appear to be a difference in the biases between the normal and VD galaxies.  Based on Figure \ref{Fig:Inc_Rot_Bias}, it is clear that \barolo\ tends to overestimate the inclination while \fat\ underestimates the inclination.  As a result, the \barolo\ fits are biased towards lower rotation curve velocities, while \fat\ is biased towards higher velocities.  \wkapp, owing to the averaging procedure, both has fewer curves to show, and does not appear to biased either towards high or low rotation velocities.  In effect, the requirement that \barolo\ and \fat\ agree with each other filters the two catalogues to produce a much more reliable catalogue with fewer contaminants and less potential biases (this is also seen in Figures \ref{Fig:LowVel_MCGResults}-\ref{Fig:UDG_MCGResults}).  Moreover, as shown in \citet{Deg2022}, the \wkapp\ uncertainties provide a better representation of the underlying variance of the models.  Ultimately, in all three panels, these contaminants fail to match the initial \mcgsuite\ models because they get the inclination wrong. 

In summary, it is indeed possible to reliably model marginally resolved galaxies with low $S/N$ and low rotation velocities at high inclinations.  This is all the more impressive given the few channels involved in some of the detections ($\sim5$ channels in some cases after the spectral smoothing).  Nonetheless, one must be cautious when considering moderate and lower inclinations.  Requiring that two different codes agree clearly provides a much more reliable catalogue, but some contaminants still remain, which are typically caused by errors in the inclinations estimates. Thus, to reliably model such low velocity systems in large surveys, great care must be taken in obtaining the inclination.  Requiring that kinematic and other inclination measures, such as the optical inclination, agree will provide a much more robust catalogue for the study of the resolved velocity function.

We have specifically tested mock WALLABY-like low rotation galaxies in this section, but the framework that we have developed for testing kinematic modelling codes in the low velocity regime is generally applicable to any survey and code combination.  It is critical to test codes extensively when applied to new regions of parameter space in order to understand their limitations.  This is particularly important as the low velocity regime is where many of the current challenges to lambda CDM cosmology lie, including the tension in the velocity function, as well as claims of galaxies with particularly low amounts of dark matter (see \citet{ManceraPina2019,ManceraPina2020} for some examples).

\section{Characterizing the Low Rotation Galaxies}\label{sec:Characterization}

\begin{figure*}
\centering
    \includegraphics[width=0.9\textwidth]{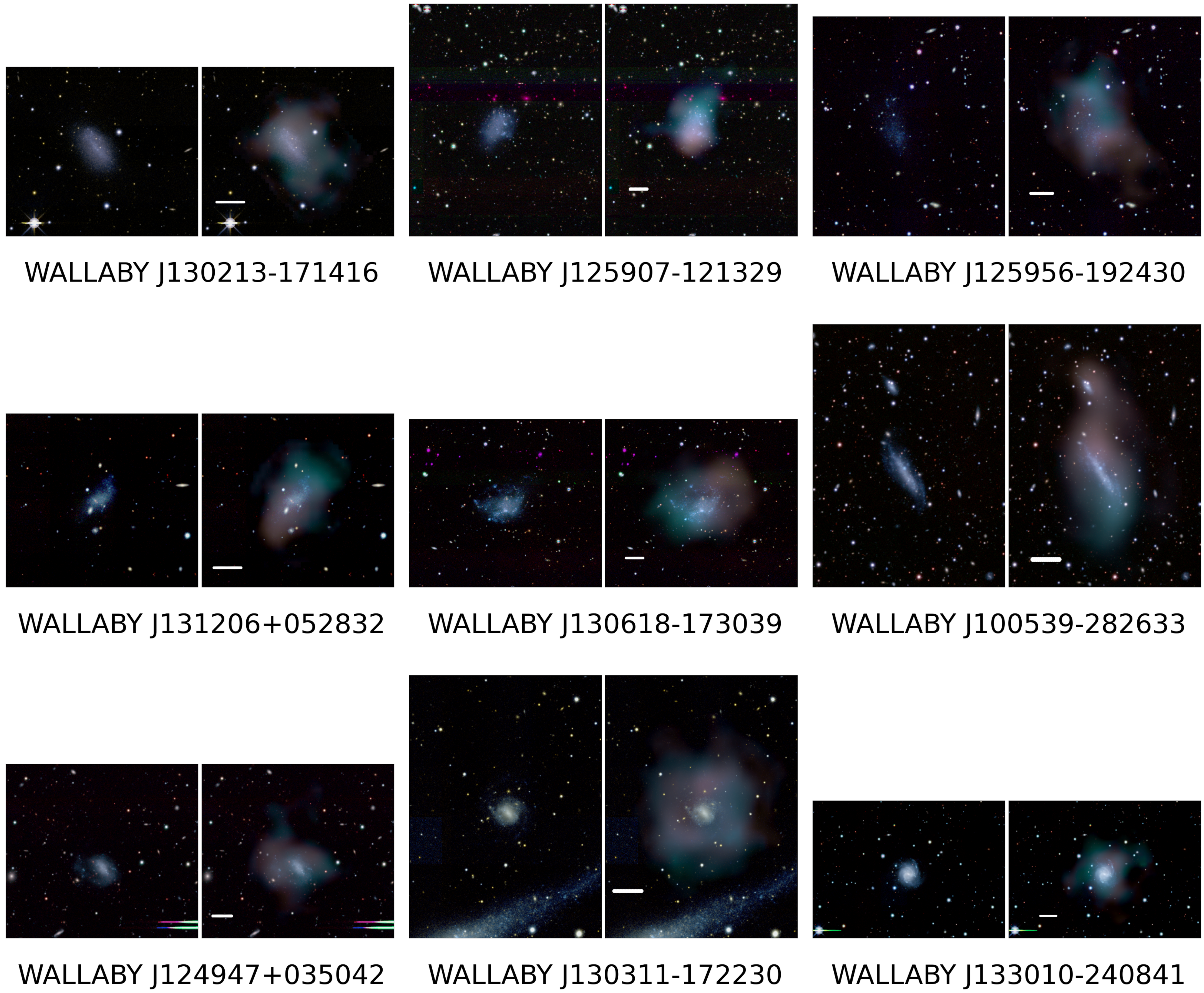} 
    \caption{Stellar  and gas overlay image image pairs (left and right of each pair respectively) for the nine low velocity galaxies with optical counterparts.  For most stellar images the blue, green, and red colors are from DECaLS $g,r,z$ band images.  In the cases of WALLABY J130213-171416 and WALLABY J130311-172230 where the $r$ band images are not available, the green color is generated by an average $(g+z)/2$ image.  The raw images are processed with CARTA to have a logarithmic stretch and adjustements to the bias and contrast, then passed through GIMP for coloring and further balancing to make the final 3-color stellar images. In the gas overlay panels, the green-red colors (approaching-receding) are set by the moment 1 map with the intensity set by the moment 0 map.  The color map itself is a custom \textsc{CosmosCanvas} map.  The white line shows the size of the WALLABY beam. }
  \label{Fig:OpticalOverlays}
\end{figure*}

With the idealized tests showing that it is indeed possible for \wkapp\ to model low velocity galaxies, we turn to our sample of eleven models with $V_{\rm{max}}\le 50~\kms$. As shown in Figures \ref{Fig:LowVel_MCGResults} and \ref{Fig:UDG_MCGResults}, the majority of our sample lie in the region that our idealized tests suggest are reliable, but nature is far more complex.  To illustrate this, Figure \ref{Fig:OpticalOverlays} shows the \hi\ gas overlaid on three-color images for the nine galaxies with DECaLS imaging (Appendix \ref{app:Obs} contains simpler overlays for the remaining two galaxies).  These overlays illustrate the complexity of these galaxies and provide key insights into understanding the reliability of the \wkapp\ models.

\begin{figure*}
\centering
    \includegraphics[width=0.9\textwidth]{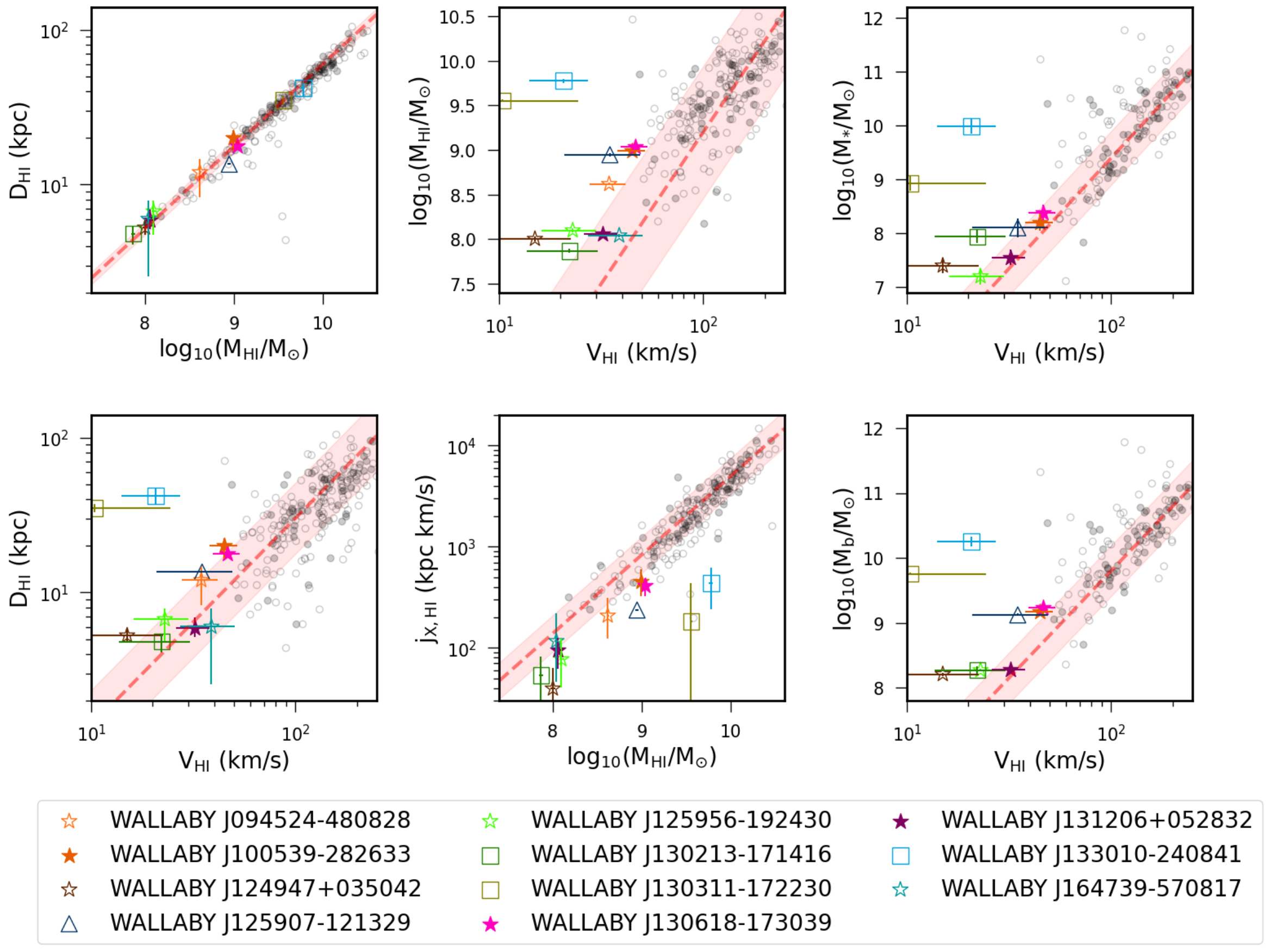} 
    \caption{The structural parameters of the 11 low-velocity galaxies compared to the larger WALLABY sample (grey circles).  The solid symbols indicate that the optical and kinematic inclinations agree.  Detections with open symbols either are missing optical inclinations (see \citet{Deg2024} for details) or the optical and kinematic inclinations disagree with each other. Squares show detections with kinematic inclinations below $40^{\circ}$, while stars show those with inclinations above $40^{\circ}$.  WALLABY J125907-121329 is indicated by a triangle even though it has an inclination of $41^{\circ}$ as it does not have robust uncertainties measured for its structural parameters (see \citealt{Deg2024} for details on the uncertainty derivations).  The red lines and shaded regions are the best fit and $1-\sigma$ widths derived from the full set of WALLABY PDR galaxies in \citet{Deg2024}.  }
  \label{Fig:LowVelScalingRelations}
\end{figure*}

In addition to examining the overlays, we can also examine the structural parameters of these galaxies in the context of scaling relations.  \citet{Deg2024} derived the \hi\ mass, size \hi\ mass, their size ($D_{\hi}$, which is the diameter where the surface density equals $1~\Msol ~ \rm{pc}^{-2}$), velocity ($V_{\hi}$, which is the velocity at $R_{\hi}=D_{\hi}/2$), and a measure of specific angular momentum ($j_{\rm{X,\hi}}=D_{\hi}/2\times V_{\hi}$) for all eleven of these galaxies.  For those with optical counterparts, \citet{Deg2024} also provides $M_{*}$ and the optical inclination using the procedure described in \cite{Arora2023} and \cite{Arora2021}.  Figure \ref{Fig:LowVelScalingRelations} shows the subset of low velocity galaxies on the size-mass (upper left), size-velocity (lower left), mass-velocity (upper center) and angular momentum-mass (lower center) relations as well as the  Tully Fisher Relation (upper right), and baryonic Tully Fisher Relation (lower right).  All the low velocity galaxies are consistent with the size-mass relation (see \citealt{Wang2016}). This is unsurprising given that the size-mass relation appears to be fundamental.  Moreover, errors in distance simply move points along the relation.

Moving to the other relations, there are some important trends to note.  We have adopted the approximate limit of $40^{\circ}$ for model reliability determined in Sec. \ref{Sec:LowVelModellingTests}. Those galaxies with inclinations above $40^{\circ}$ (indicated by stars) generally agree with the underlying scaling relations within their uncertainties, while the majority of those below this limit (indicated by squares) disagree with the relations. Moreover, the three galaxies with consistent kinematic and photometric inclinations that are also above $40^{\circ}$ lie closest to the rest of the scaling relations.  

Going through the galaxies on an individual basis, there are some surprises.  The three galaxies in the top row of Figure \ref{Fig:OpticalOverlays} all appear to be `normal' rotating disks, but they all have inconsistent kinematic and photometric inclinations.  In the case of WALLABY J130213-171416 (upper left in Figure \ref{Fig:OpticalOverlays}), the kinematic inclination is $28.7^{\circ}$, which lies well below the threshold of reliability.  WALLABY 125907-121329 does have a kinematic inclination of $41^{\circ}\pm4^{\circ}$, which would place it at the edge of reliability.  However, \citet{Deg2024}, were unable to measure an upper limit on $D_{\hi}$ meaning that the uncertainties on the structural parameters are not robust.  Thus, it is unclear whether it is reliable or not.  WALLABY J125956-192430 is interesting owing to its low surface brightness. As with the other two galaxies, the photometric and kinematic inclinations are inconsistent.  But, unlike the other two, the low surface brightnesses of WALLABY J125956-192430 means that the photometric inclination tends to be unreliable due to the stochastic nature of star formation.  Dudley et al. (submitted) examines this galaxy in much greater detail and was able to generate a greatly improved kinematic model of the galaxy using a customized approach that considers both the photometric and \hi\ distributions.  They find that WALLABY J125956-192430 is an UDG at a moderate inclination with central surface brightness of $\mu_{0,g}=24.55$ mag arcsec$^{-2}$, a half-light radius of $R_{1/2}=1.27$ kpc (see Dudley et al. (submitted) for the derivation of these quantities, as well as a study of its DM content).

The central row of Figure \ref{Fig:OpticalOverlays} consists of galaxies that appear to be disturbed in some fashion.  WALLABY J13120-052832 shows signs of an interaction and WALLABY J130618-173039 has an irregular morphology.  Most intriguing is WALLABY J100539-282633, which consists of two different galaxies.  We have retained this galaxy in our sample as our selection and the underlying kinematic model are based purely on the \hi\ observation.  It is important to note here that the \hi\ mass and inclination are derived from the total system, while the photometric inclination and stellar mass are derived from the primary only.  Despite their complexity, all three galaxies are have consistent photometric and kinematic inclinations and all three consistent with the larger WALLABY scaling relations when considering their uncertainties.  For the pair of galaxies, this is likely due to the dominance of the primary in terms of mass and kinematics.  It is likely that a more careful mask would further improve the fit.  For the other two, the kinematic models are likely successful as the disturbances are primarily in the outer regions.  Since both \barolo\ and \fat\ fit iteratively from the inside out, those disturbances may not affect the core rotating disks.

The bottom row of Figure \ref{Fig:OpticalOverlays} shows that WALLABY J124947+035042, WALLABY J130311-172230, and WALLABY J133010-24084 are in fact face-on spirals.  Two of these have kinematic inclinations well below the $40^{\circ}$ reliability limit, and are the largest outliers in the scaling relations.  WALLABY J124947+035042 is the largest outlier among galaxies with inclinations above $40^{\circ}$, but it's inclination of $45^{\circ}\pm5^{\circ}$ is consistent with the reliability limit.  This suggests that galaxies that have lower limits to their inclinations should be treated with caution (which would also include WALLABY 125907-121329, which has an inclination of $41^{\circ}\pm 4^{\circ}$).

In the discussion of Figure \ref{Fig:LowVel_RCs} we noted that some galaxies appeared to have falling rotation curves.  All those galaxies also have inclinations below our $40^{\circ}$ limit.  Thus, while falling rotation curves are interesting, none of the falling rotation curves in our sample lie in a regime where the models are reliable.  It will be more useful to explore whether there are falling rotation curves in the larger sample of WALLABY kinematic models to see if any are present and the implications that such rotation curves would have on their structure.

In summary, despite their complexity, those galaxies above the $40^{\circ}$ inclination limit of Sec. \ref{Sec:LowVelModellingTests} lie within $\approx1-2~\sigma$ of scaling relations defined by the larger WALLABY population.  While our tests are specific to WALLABY-like galaxies modelled using a particular set of algorithms, it is reasonable to be cautious of any model with kinematic inclinations below $40^{\circ}$.  In general, confirming models at these lower inclinations will require extensive testing.  Additionally, it is useful to have multiple, independent inclination measurements that consistent with each other.  While gas and stars may be misaligned in nature, such systems will not have simple, rotating disks. 

It is worth noting here that these results generally agree with \citet{Lelli2024}. They examined a suite of gas rich UDGs, including a subset that appear to fall off the bTFR \citep{ManceraPina2019,ManceraPina2020,Shi2021}. \citet{Lelli2024} noted that gas rich UDGs and other LSBs tend to scatter in the same direction on the bTFR.  These typically are at moderate to low inclinations, which, when the inclination is overestimated, will lead to an underestimate of the rotation velocity. It is also worth noting that, for many of the galaxies in the \citet{ManceraPina2020} sample, the inclination is measured by either the optical or \hi\ morphology rather than the kinematic inclination.  Nonetheless, both the \citet{Lelli2024} results as well as our study highlight the great challenge faced when modelling low-velocity galaxies in the low resolution, low $S/N$ regime.  The great advantage of WALLABY is the large number of detections that will be generated by the full survey, allowing for a more careful selection of reliable models via inclination cuts as well as optical followups of low mass populations.

\section{Discussion and Conclusions}\label{Sec:Conclusions}

A core goal of WALLABY is kinematically modelling a statistically significant sample of galaxies in the low mass/low velocity regime.  This will enable probes of the resolved velocity function, tests of DM content at the low mass end and more.  In order to explore the reliability of modelling in this regime,  we have examined kinematic models of such galaxies using both idealized tests and 11 WALLABY PDR detections with maximum velocities below $50~\kms$.

We began by testing \barolo, \fat, and \wkapp\ on 200 mock idealized galaxies with velocities ranging between $20-70~\kms$ generated via \mcgsuite.  These tests demonstrated that:
\begin{enumerate}
    \item The averaging process of \wkapp\ removed many of the false positives found in the \barolo\ and \fat\ runs. 
    \item Above $\approx 40^{\circ}$ and \ellmaj$\ge{2~\rm(beams)}$ all three codes are fairly successful in reproducing the underlying model.
    \item Below $\approx 40^{\circ}$, the false positives comprise $50\%$ of the sample for \wkapp.  For \fat\ and \barolo\ the false positive rate is significantly larger than $50\%$. 
    \item The false positive rate below $40^{\circ}$ does not appreciably change when using the smoothed $12~\kms$ or full resolution $4~\kms$ cubelets.  This highlights the fact that the driver of the false positives in these idealized cases is the inclination itself. 
    \item Generally, at low inclinations, \barolo\ is biased to measuring a higher inclination and \fat\ to a lower inclination.  This leads to \barolo\ inferring lower velocities and \fat\ to inferring higher velocities than the underlying model.
    \item Given these results, we recommend being particularly cautious when considering low rotation models with kinematic inclinations below $40^{\circ}$ as our tests have shown that fits of even idealized galaxies are unreliable in this regime across multiple codes.
\end{enumerate}
These results suggest that a key improvement in kinematic modelling low velocity galaxies with WALLABY-like resolutions and $S/N$ values is to obtain independent inclination measurements, as it is inclination errors that cause the majority of the false positives. However, even photometric inclinations are difficult to measure for low mass dwarfs.  And, the biases in the measured inclinations will lead to biases in the inferred galaxy properties \citep{Read2016}.  Given these, when a larger sample of low rotation galaxies is available, it may be wiser to simply remove galaxies with low kinematic inclinations. The idea of removing low inclination galaxies from a sample is not new \citep{Rhee2004,Read2016}.  It has been discussed in the context $H\alpha$ observations \citep{Rhee2004}, IFU observations \citep{KuziodeNaray2006}, and other tracers.  Here we have focused on quantifying the limit of reliability for kinematic modelling in the low resolution, low $S/$, low rotation regime for \hi\ observations.

It is worth discussing point 4 in greater depth. As noted in Sec. \ref{Sec:LowVelModellingTests}, the low rotation regime is perhaps the most challenging for kinematic modelling.  Non-circular motions, various disruptions to the morphology, and even relatively small velocity dispersions will affect fitting codes and may lead to biases in the morphology.  Thus it is critical to extensively test codes in this regime.  While our tests are specific to the WALLABY context, the framework we have developed can be applied to any code and untargetted \hi\ survey.  It is always possible to build idealized mock cubes (whether via \mcgsuite\ or some other tool) from low rotation models and compare the results of a particular fitting code to the underlying models. 

Testing in the low velocity regime using idealized galaxies does present a rather optimistic scenario. There are many features that such models will not capture such as holes, bubbles, bars, inflows/outflows, and more.  Each of these will cause further challenges to fitting that will be important to consider.  In the future, these tests can be improved upon via the use of mock observations of hydrodynamical simulations.  However, these pose their own unique challenges (see for example \citealt{Perron-Cormier2025}) due to difficulties in making reliable mock observations as well as the necessity of testing the particular simulation itself.  However, it is also important to note the scale of such features and the radial regimes where they may affect the kinematic models.  For example, in the marginally resolved regime a bar, if present, would affect the innermost region (which will already be affected by beam smearing), and may not affect the global rotation measure required for a probe of the bTFR.  

The importance of this testing can be seen by contrasting the results of Sec. \ref{sec:Characterization} to the work of \citet{Deg2022} and \citet{Deg2024}.  In \citet{Deg2022}, \wkapp\ was tested extensively for its applicability to WALLABY, but it was not tested in the low rotation regime.  In Sec. \ref{sec:Characterization} we demonstrated that three of these low rotation models are actually face-on spirals, one of them is actually a pair of galaxies, and a number of others have unreliable inclinations.  The few detections with consistent optical and kinematic inclinations lie on or very near all the WALLABY scaling relations explored in \citet{Deg2024}.  The others appear to be outliers from the scaling relations due purely to unreliability of their models (in particular due to unreliable inclinations in the case of the three face-on spirals) rather than differences in their underlying structure.

It is worth comparing our results to those of \citet{Siljeg2024}, who examined a set of 24 dwarfs and UDGs.  They found significant discrepancies from the bTFR, even for high inclination galaxies. These galaxies fall off the bTFR in a similar fashion to galaxies in our sample with inconsistent optical and kinematic inclinations.  This is particularly notable as the \citet{Siljeg2024} sample tend to have disagreeing optical and \hi\ inclinations (which were derived from the morphology rather than kinematic modelling).  This again highlights the challenge of measuring inclinations for low mass galaxies where stochastic star formation, warps, and other disruptions can strongly affect the measurement.

Continuing with this theme, we also consider the results of \citet{Obreschkow2013b}.  They compared estimates of the bTFR using standard methodologies involving inclination corrected profile widths. They found that using an `inclination-free' maximum likelihood estimation worked better than methods than using an inclination corrected method when inclination uncertainties are $\ge10^{\circ}$. Our situation is slightly different in that we find that models with kinematic inclinations above $40^{\circ}$ are generally reliable.  In cases where they also have consistent photometric and kinematic inclinations the models are significantly more reliable, leading to less biases in the results.  This is seen in the closer agreement of our sample of galaxies with consistent inclinations with the larger WALLABY population in terms of their structural parameters.

Kinematically modelling galaxies with low rotation velocities is critical to WALLABY.  Understanding this population of galaxies is critical for understanding dark matter, the bTFR, measuring the resolve velocity function and more.  We have shown that low rotation WALLABY detections can be modelled using a number of different codes \textit{when the kinematic inclination is above $40^{\circ}$.}   This is true even for galaxies that show disturbed morphologies, as the inner regions can often still be characterized by a rotating disk.

One particular advantage of the TR modelling approach is that it can be applied to large cosmological simulations as well as observations.  Tools like \textsc{MARTINI} \citep{Oman2019a,Oman2019b,Oman2024} and other codes \citep{Read2016,Guglielmetti2023,Perron-Cormier2025} are capable of building realistic mock \hi\ cubes that can be analyzed with the exact same modelling codes as observations, allowing for a detailed statistical comparison of observations and simulations that has heretofore been impossible.  It is only WALLABY that will have the large sample size necessary to resolve the tensions in the velocity function.  And WALLABY will also cover a large enough sample of different environments, masses, and properties that it has a relatively straightforward selection function.  Already first comparisons are being made to mock observations using morphometrics \citep{Perron-Cormier2025}.  As the full survey progresses, using the lessons from this study of low velocity WALLABY detections, the prospects for measuring the velocity function and resolving tensions between observations and simulations are better than ever.  

\section*{Data Availability}

All WALLABY data used in this analysis is available through links on the  \href{https://wallaby-survey.org/data/data-pilot-survey-dr1/}{WALLABY data page}.  DOIs for the PDR1 and PDR2 source files are \href{https://doi.org/10.25919/09yg-d529}{https://doi.org/10.25919/09yg-d529} and \href{https://doi.org/10.25919/qw7w-tn96}{https://doi.org/10.25919/qw7w-tn96} respectively, and the DOIS for the PDR1 and PDR2 kinematic models are \href{https://doi.org/10.25919/213m-p819}{https://doi.org/10.25919/213m-p819} and \href{https://doi.org/10.25919/7w8n-9h19}{https://doi.org/10.25919/7w8n-9h19} respectively.  The mock cube generator and kinematic modelling codes are available via GitHub at \href{https://github.com/CIRADA-Tools/WKAPP}{\wkapp} and \href{https://github.com/CIRADA-Tools/MCGSuite}{\mcgsuite} respectively.  All other products and analysis scripts are available upon request. 

\section*{Acknowledgements}

Thanks to the referee for the useful comments and suggestions for improvements.  And thanks to A. Bosma and P. Mancera Pi\~na for helpful discussions and comments.

This scientific work uses data obtained from Inyarrimanha Ilgari Bundara / the Murchison Radio-astronomy Observatory. We acknowledge the Wajarri Yamaji People as the Traditional Owners and native title holders of the Observatory site. CSIRO’s ASKAP radio telescope is part of the Australia Telescope National Facility (https://ror.org/05qajvd42). Operation of ASKAP is funded by the Australian Government with support from the National Collaborative Research Infrastructure Strategy. ASKAP uses the resources of the Pawsey Supercomputing Research Centre. Establishment of ASKAP, Inyarrimanha Ilgari Bundara, the CSIRO Murchison Radio-astronomy Observatory and the Pawsey Supercomputing Research Centre are initiatives of the Australian Government, with support from the Government of Western Australia and the Science and Industry Endowment Fund.

This work was facilitated by the Australian SKA Regional Centre (AusSRC), Australia’s portion of the international SKA Regional Centre Network (SRCNet), funded by the Australian Government through the Department of Industry, Science, and Resources (DISR; grant SKARC000001). AusSRC is an equal collaboration between CSIRO – Australia’s national science agency, Curtin University, the Pawsey Supercomputing Research Centre, and the University of Western Australia.

This research was undertaken thanks in part to funding from the Canada First Research Excellence Fund through the Arthur B. McDonald Canadian Astroparticle Physics Research Institute.

The Legacy Surveys consist of three individual and complementary projects: the Dark Energy Camera Legacy Survey (DECaLS; Proposal ID \#2014B-0404; PIs: David Schlegel and Arjun Dey), the Beijing-Arizona Sky Survey (BASS; NOAO Prop. ID \#2015A-0801; PIs: Zhou Xu and Xiaohui Fan), and the Mayall z-band Legacy Survey (MzLS; Prop. ID \#2016A-0453; PI: Arjun Dey). DECaLS, BASS and MzLS together include data obtained, respectively, at the Blanco telescope, Cerro Tololo Inter-American Observatory, NSF’s NOIRLab; the Bok telescope, Steward Observatory, University of Arizona; and the Mayall telescope, Kitt Peak National Observatory, NOIRLab. Pipeline processing and analyses of the data were supported by NOIRLab and the Lawrence Berkeley National Laboratory (LBNL). The Legacy Surveys project is honored to be permitted to conduct astronomical research on Iolkam Du’ag (Kitt Peak), a mountain with particular significance to the Tohono O’odham Nation.

NOIRLab is operated by the Association of Universities for Research in Astronomy (AURA) under a cooperative agreement with the National Science Foundation. LBNL is managed by the Regents of the University of California under contract to the U.S. Department of Energy.

This project used data obtained with the Dark Energy Camera (DECam), which was constructed by the Dark Energy Survey (DES) collaboration. Funding for the DES Projects has been provided by the U.S. Department of Energy, the U.S. National Science Foundation, the Ministry of Science and Education of Spain, the Science and Technology Facilities Council of the United Kingdom, the Higher Education Funding Council for England, the National Center for Supercomputing Applications at the University of Illinois at Urbana-Champaign, the Kavli Institute of Cosmological Physics at the University of Chicago, Center for Cosmology and Astro-Particle Physics at the Ohio State University, the Mitchell Institute for Fundamental Physics and Astronomy at Texas A\&M University, Financiadora de Estudos e Projetos, Fundacao Carlos Chagas Filho de Amparo, Financiadora de Estudos e Projetos, Fundacao Carlos Chagas Filho de Amparo a Pesquisa do Estado do Rio de Janeiro, Conselho Nacional de Desenvolvimento Cientifico e Tecnologico and the Ministerio da Ciencia, Tecnologia e Inovacao, the Deutsche Forschungsgemeinschaft and the Collaborating Institutions in the Dark Energy Survey. The Collaborating Institutions are Argonne National Laboratory, the University of California at Santa Cruz, the University of Cambridge, Centro de Investigaciones Energeticas, Medioambientales y Tecnologicas-Madrid, the University of Chicago, University College London, the DES-Brazil Consortium, the University of Edinburgh, the Eidgenossische Technische Hochschule (ETH) Zurich, Fermi National Accelerator Laboratory, the University of Illinois at Urbana-Champaign, the Institut de Ciencies de l’Espai (IEEC/CSIC), the Institut de Fisica d’Altes Energies, Lawrence Berkeley National Laboratory, the Ludwig Maximilians Universitat Munchen and the associated Excellence Cluster Universe, the University of Michigan, NSF’s NOIRLab, the University of Nottingham, the Ohio State University, the University of Pennsylvania, the University of Portsmouth, SLAC National Accelerator Laboratory, Stanford University, the University of Sussex, and Texas A\&M University.

BASS is a key project of the Telescope Access Program (TAP), which has been funded by the National Astronomical Observatories of China, the Chinese Academy of Sciences (the Strategic Priority Research Program “The Emergence of Cosmological Structures” Grant \# XDB09000000), and the Special Fund for Astronomy from the Ministry of Finance. The BASS is also supported by the External Cooperation Program of Chinese Academy of Sciences (Grant \# 114A11KYSB20160057), and Chinese National Natural Science Foundation (Grant \# 12120101003, \# 11433005).

The Legacy Survey team makes use of data products from the Near-Earth Object Wide-field Infrared Survey Explorer (NEOWISE), which is a project of the Jet Propulsion Laboratory/California Institute of Technology. NEOWISE is funded by the National Aeronautics and Space Administration.

The Legacy Surveys imaging of the DESI footprint is supported by the Director, Office of Science, Office of High Energy Physics of the U.S. Department of Energy under Contract No. DE-AC02-05CH1123, by the National Energy Research Scientific Computing Center, a DOE Office of Science User Facility under the same contract; and by the U.S. National Science Foundation, Division of Astronomical Sciences under Contract No. AST-0950945 to NOAO.

\bibliography{LowVelCharacteristics}{}
\bibliographystyle{aasjournal}

\appendix

\section{MCGSuite}\label{app:MCGSuite}

\mcgsuite\ is a tool for generating idealized realistic mock observations using scaling relations and TR models.  A core advantage of \mcgsuite\ is that it allows a user to input a set of observational quantities (\hi\ mass, diameter in beams, and observed geometry) and it will generate a realistic mock datacube. In detail, \mcgsuite\ consists of a core TR generating code, \texttt{MockCubeGenerator}, written in Fortran.  This is a very flexible code, allowing for a user to generate a mock observation for an arbitrary TR model.  The user must specify the center point, systemic velocity, rotation velocity, radial velocity, vertical velocity, velocity dispersion, surface density, thickness, position angle, and inclination of each ring.  This allows for the generation of complicated models, including polar rings and warps. In addition, the user must also specify the properties of the data cube (pixel size/channel width, number of pixels/channels, center point, beam size), as well as the noise level.  With all parameters supplied, \texttt{MockCubeGenerator} will generate an unconvolved noiseless cube, a convolved noiseless cube, and a noisy convolved cube. 

The process of generating a mock cube is as follows:
\begin{enumerate}
    \item Initialize the rings
    \item Fill each ring with a set of tracer particles.  The particles are initialized in local coordinates (i.e. the $x-y$ plane is the plane of the disk galaxy, but calculate in arcseconds).  The particle number density is set to be uniform in $(R,~\theta,~z)$ coordinates, while the flux is set by the surface density of the particular ring.  The particle velocities in these local coordinates are set by the rotation, radial velocity, vertical velocity, and velocity dispersion parameters.
    \item The tracer particles are rotated into observed coordinates via the position angle and inclination parameters, as well as shifted to the systemic velocity.
    \item Once the full particle array for the full set of rings are determined, they are placed into the datacube via simple summation.
    \item The datacube is then convolved by the beam using the \textsc{fftw} library \citep{FFTW2005}.
    \item An empty cube filled with Gaussian noise is then convolved by the same beam such that the convolved noise cube matches the input noise parameter.  This convolved noise cube is added to the convolved noiseless cube to generate the final mock cube.
\end{enumerate}

The \texttt{MockCubeGenerator} code can generate a wide variety of complex models owing to the flexibility of TR models.  However, TR models may not be particularly realistic.  And, due to the large number of free parameters, \texttt{MockCubeGenerator} can be rather unwieldy to use.  To solve these problems, \textsc{MCGSuite} includes three different python wrappers that exploit scaling relations to allow users to easily generate sets of realistic, idealized galaxies.  They allow the user to specify a size in beams, an \hi\ mass, and an inclination and position angle which are coupled with a supplied beam size, pixel size, channel size, and noise level to generate a mock cube (or set of mock cubes).

The first wrapper is \texttt{make\_galaxy\_MCG}, which generates a single model.  The second is \texttt{make\_suite\_MCG}, which generates a suite of models along a grid of sizes, masses, inclinations, and position angles.  It will also generate multiple realizations of the same underlying model, which enables tests of the uncertainty due to the underlying noise.  The third wrapper is \texttt{make\_suite\_MCG\_random} which generates an arbitrary number of cubes from a random distribution of parameter values.  This can be used to create a suite of mock cubes that will fully sample some parameter space.

The key aspect of all three wrappers is the generation of TR models from simple parameters via scaling relations.  The first relation is the \hi\ size-mass relation of \citet{Wang2016}.  This relation is 
\begin{equation}\label{Eq:SizeMass}
    \log_{10}(D_{\hi})=(0.506\pm0.003)\log_{10}(M_{\hi})+(3.293\pm0.009)~,
\end{equation}
where $M_{\hi}$ is the total \hi\ mass and $D_{\hi}$ is the diameter in kpc where the \hi\ surface density reaches $1~\Msol~\rm{pc}^{2}$.  The \hi\ size-mass relation is among the tightest scaling relations in all astrophysics, with galaxies of almost all types falling on this relation.  In \textsc{MCGSuite} the input \hi\ mass is converted to $R_{\hi}=0.5 D_{\hi}$ via Eq. \ref{Eq:SizeMass} ignoring the variance measured in \citet{Wang2016} (that is only using the 0.506 and 3.293 factors).

The structural velocity, $V_{\hi}$, which is the velocity at $R_{\hi}$, is determined from the \hi\ mass and \hi\ velocity functions from the ALFALFA survey \citep{Giovanelli05,Haynes18}.  In detail, the Schechter function to the \hi\ mass function of \citet{Martin10} and the modified Schechter function fit to the inclination-corrected velocity function of \citet{Papastergis15}, which are both derived from the ALFALFA 40\% catalogue \citep{Haynes11}, are combined together via abundance matching to determine a relationship between $M_{\hi}$ and $V_{\hi}$.   We then use a fifth-order polynomial fit to this relationship to compute $V_{\hi}$ from $M_{\hi}$:
\begin{equation}
\log(\VHI~\kms) =  0.0012992\xHI^5 -0.0620705\xHI^4  + 1.210101\xHI^3 -11.98959\xHI^2  + 60.46594\xHI -122.66267~,
\label{eq:MHIVHI}
\end{equation}
where $\xHI = \log(M_{\hi}/\Msol)$.  

Once $R_{\hi}$ and $V_{\hi}$ are calculated, the overall rotation curve is calculated using the Polyex function \citep{Giovanelli02}:
\begin{equation}\label{Eq:Polyex}
    V_{\rm{rot}}(R)=V_{\rm{PE}}  \left(1-e^{(-R/R_{PE})}\right)\left(1+a_{PE}*R/R_{PE}\right)~,
\end{equation}
where $R_{PE}$ is the turnover radius, $V_{PE}$ is the amplitude, and $a_{PE}$ is the outer slope. These constants are determined for each $M_{\hi},~V_{\hi}$ pair via empircal relations.  Firstly, the inner rotation curve shape is designed to follow the template relations derived by \citet{Catinella06} from $\sim2200$ long-slit H$\alpha$ spectra of nearby disk galaxies. \citet{Catinella06} parameterizes their relation in terms of the optical radius \Ropt, where they fit linear relations to \rPE/\Ropt and \aPE\ and an exponential + linear relation to \VPE\ reported in their table 2 as a function of absolute magnitude \MI.  We find the \MI\ for which the  best-fitting \rPE, \aPE\ and \VPE\ reproduce our initial guess for the rotation curve amplitude at \Ropt\ of \Vrot(\Ropt) = 0.95\VHI.  Secondly, we aim to have our outer rotation curve slopes follow the trend of SPARC galaxies with $R_{\hi}$ \citep{Lelli16a}.  To that end, \mcgsuite\ retains \rPE\ (converted from \rPE/\Ropt\ assuming \Ropt = \RHI/1.7 as in \citealt{Broeils97}) and uses it to find the \aPE\ that reproduces the logarithmic outer slope from $0.5\RHI < r < \RHI$ corresponding to the model galaxy \RHI\ from the SPARC scaling relation derived by \cite{Dutton19}. Finally, we re-normalize \VPE\ using \rPE\ and \aPE\ determined as described above such that \Vrot(\RHI) = \VHI\ as required.  

The overall surface density profile of the model is a combined Gaussian plus exponential profile given by
\begin{equation}
    \frac{\Sigma_{\hi}(R)}{\Sigma_{\textrm{max}}} = \exp \left(- \frac{(R-{R_{\Sigma m}})^2}{2\sigma_{\Sigma}^2} \right) + \left( 1 - \sqrt{\frac{V_{\hi}}{V_{\Sigma}}} \right) \exp^{-R/R_{\Sigma S}} ~,
    \label{eq:sigHI}
\end{equation}
where $\Sigma_{\textrm{max}}$ is the maximum surface density, $\sigma_{\Sigma}$ and $R_{\Sigma S}$ are scale radii for the Gaussian and exponential respectively, $R_{\Sigma m}$ is the radius where the Gaussian term peaks, and $V_{\Sigma}$ is a scale velocity.  The idea behind this profile is that the Gaussian term dominates at $R>0.5R_{\hi}$ and the exponential modifies the profile at $R<0.5 R_{\hi}$.  In low mass systems with $V_{\hi}<V_{\Sigma}$ the exponential adds mass to the center, and in high mass systems it removes mass from the center.  We set $R_{\Sigma S} = 0.2\RHI$ and adopt $V_{\Sigma} = 120\,\kms$ as the threshold between low- and high-mass systems \citep[e.g.][]{Dalcanton04,Kannappan13}.  For the Gaussian, we set $R_{\Sigma m}=0.4\RHI$ based on \citet{Martinsson16} and then fix $\Sigma_{\rm{max}}$ such that $\Sigma_{\hi}(R_{\hi})=1~\Msol~\textrm{pc}^{-2}$.  Finally, we set the Gaussian scale radius to $\sigma_{\Sigma}=0.33\RHI$ (close to the value measured by \citealt{Martinsson16} of $\sigma_{\Sigma} = 0.34\RHI$) which ensures that 85\% of the model \hi\ mass is located within \RHI, as measured by \citet{Wang2016}.  We also set the scale height of assuming that it is a thindisk assuming a constant velocity dispersion $\sigma_{\hi} = 10\,\kms$, consistent with recent measurements near the disk edge from high-resolution \hi\ maps \citep{Tamburro09}.

In summary, by specifying $M_{\hi}$, the surface density profile and rotation curve are specified completely by the above relations.  Then the user-specified size in beams is used to calculate the mock galaxy's convert the radii from kpc to arcseconds, as well as calculate the distance and recessional velocity via the Hubble flow.  \mcgsuite\ then takes these observed profiles coupled with the specified inclination and position angle and converts them into TR parameters for the \texttt{MockCubeGenerator} code and runs it to make a mock cubelet.

The two other Python wrappers comprising \mcgsuite\ are the \texttt{make\_suite\_MCG} and \texttt{make\_suite\_MCG\_random}.  These generalize \texttt{make\_galaxy\_MCG} to automatically generate large sets of mock observations.  The \texttt{make\_suite\_MCG\_random} will generate an arbitrary number of mock cubes with parameters drawn randomly from a user specified range of masses, sizes, inclinations, and position angles.  \texttt{make\_suite\_MCG}  on the other hand generates a suite covering a user specified grid of masses, sizes, inclinations, and position angles.  it has the option to build more than one realization per grid point in order to allow testing of how the noise affects measurements of the mock cube.

All three codes also have an option to generate VD mock galaxies.  This mode was developed as there are certain galaxy classes, like UDGs where the rotation velocities may follow different scaling relations.  Since we do not know what these relations might be, the VD models decouple the relationship between $M_{\hi}$ and $V_{\hi}$.  Instead, the user specifies $V_{\hi}$ explicitly and then \mcgsuite calculates the rotation curve using Eq. \ref{Eq:Polyex} with $V_{\rm{PE}}=275$, $R_{PE}=0.126$, and $a_{PE}=0.08$.  Since all galaxies with \hi\ disks still fall on the size-mass relation, $R_{\hi}$ and the surface density profile of the VD mocks are determined the same way as for standard galaxies.

\begin{figure*}
\centering
    \includegraphics[width=0.9\textwidth]{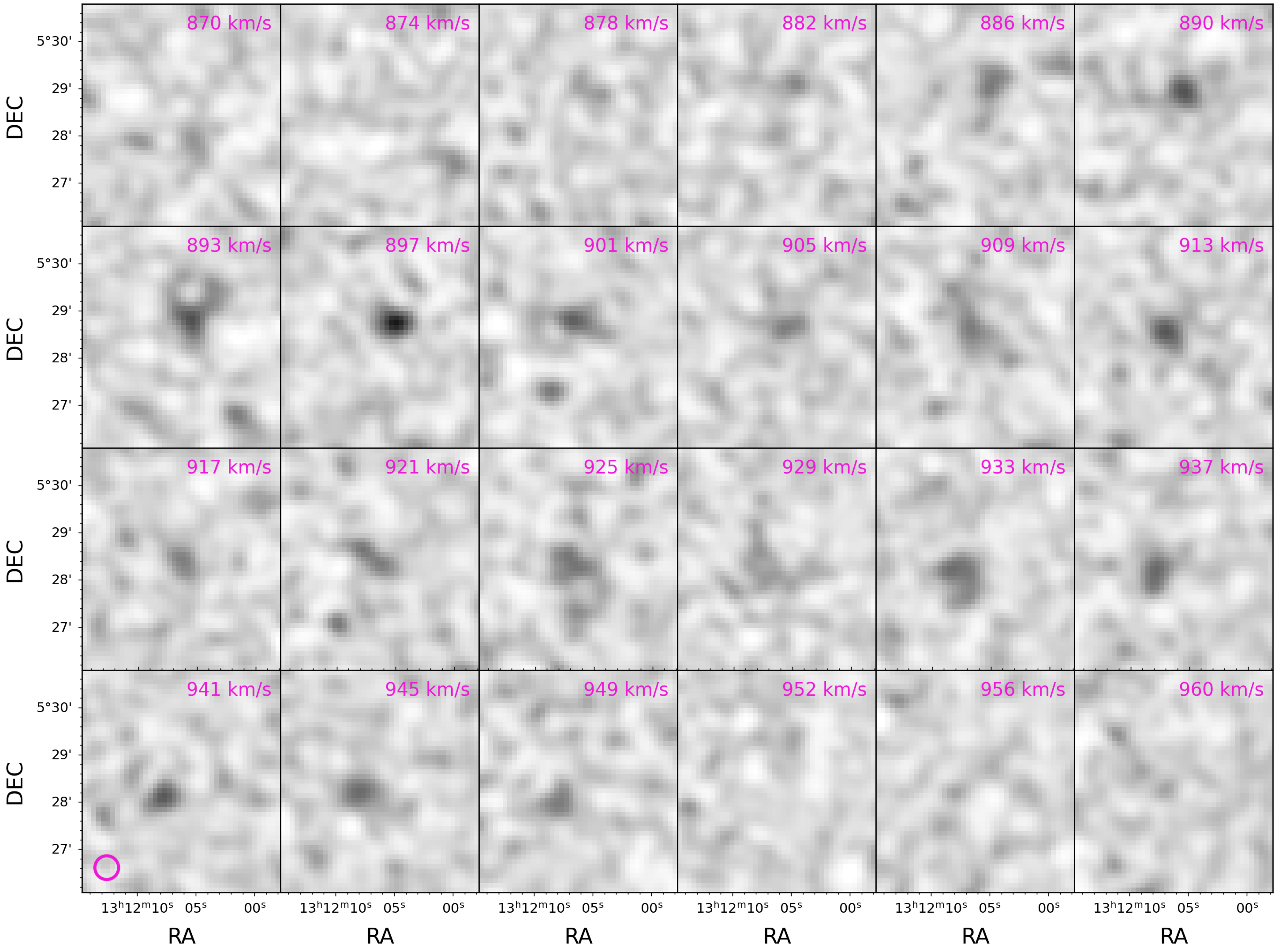} 
    \caption{The channel maps for WALLABY J131206+052832. The magenta circle in the lower left shows the size of the beam.}
  \label{Fig:Real131Maps}
\end{figure*}

\begin{figure*}
\centering
    \includegraphics[width=0.9\textwidth]{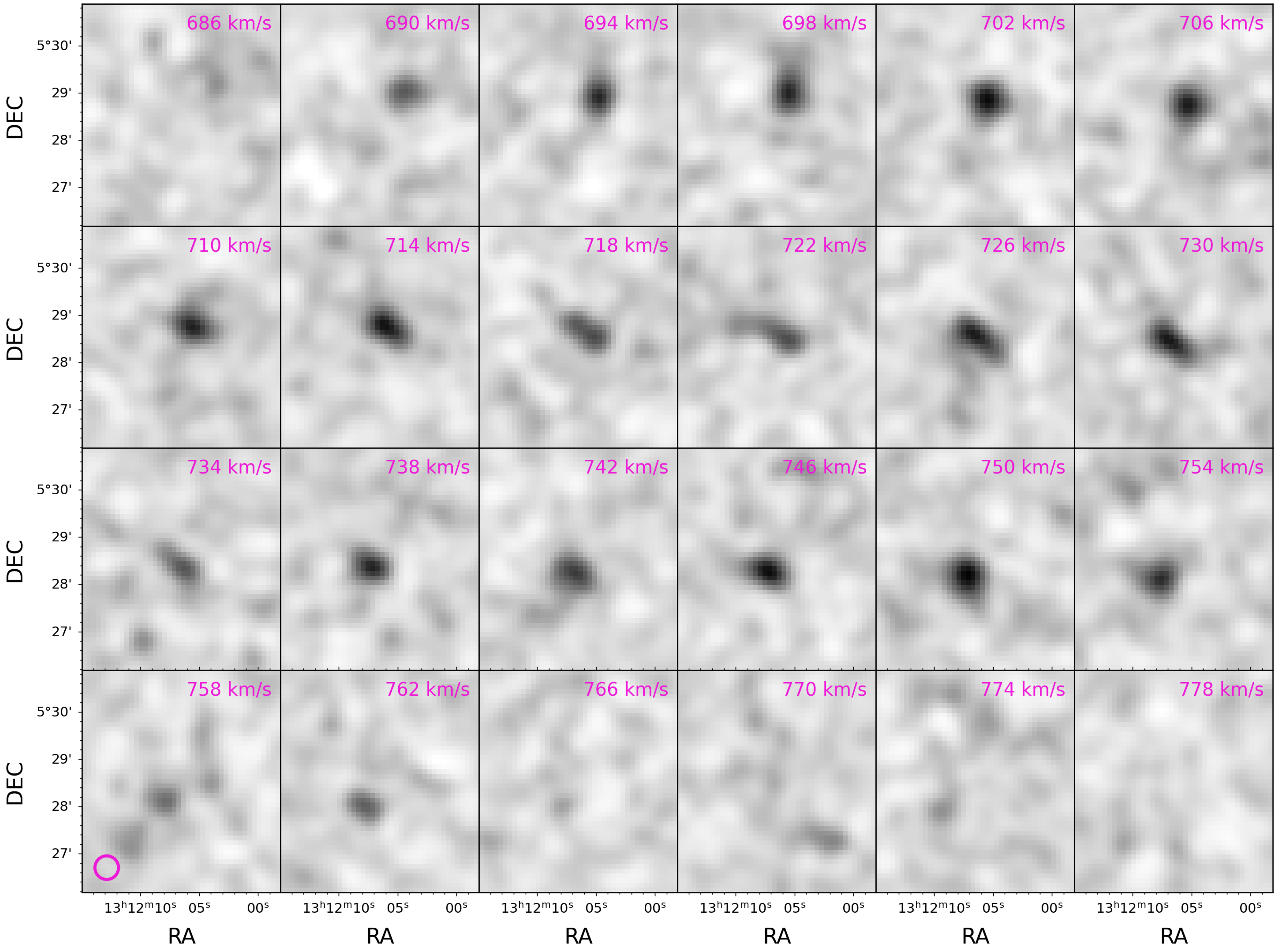} 
    \caption{The channel maps for an \mcgsuite\ mock cube made with $\log_{10}(M_{\hi}/\Msol)=8.04$, $D_{\hi}=4~\rm{(beams)}$, and $i_{\rm{kin}=56^{\circ}}$.  The mass, diameter, and inclination are set based on the measurements of WALLABY J131206+052832.  The magenta circle in the lower left shows the size of the beam.}
  \label{Fig:Mock131Maps}
\end{figure*}

As a brief demonstration of \mcgsuite\, Figure \ref{Fig:Real131Maps} shows the channel maps for WALLABY J131206+052832 while Figure \ref{Fig:Mock131Maps} shows channel maps for an \mcgsuite\ cube made with similar parameters ($\log_{10}(M_{\hi}/\Msol)=8.04$, $D_{\hi}=4~\rm{(beams)}$, $i_{\rm{kin}=56^{\circ}}$).  To be clear, the \mcgsuite\ maps shown in Figure \ref{Fig:Mock131Maps} are not meant to match those of WALLABY J131206+052832 in Figure \ref{Fig:Real131Maps}.  The \mcgsuite\ realization is an idealized version of a galaxy that perfectly follows the underlying scaling relations used to build the model.  It clear has a higher $S/N$ and a significantly more regular morphology.  It is also at a different systemic velocity due to \mcgsuite\ using the Hubble flow to calculate systemic velocities.  Nonetheless, the two cover similar velocity ranges and have similar orientations.  While this particular mock cube was designed to be similar to WALLABY J131206+052832, any number of these can be generated covering a large range of parameter space. 

\section{Testing Diagnostic Plot}\label{app:TestingPlot}

\begin{figure*}
\centering
    \includegraphics[width=0.9\textwidth]{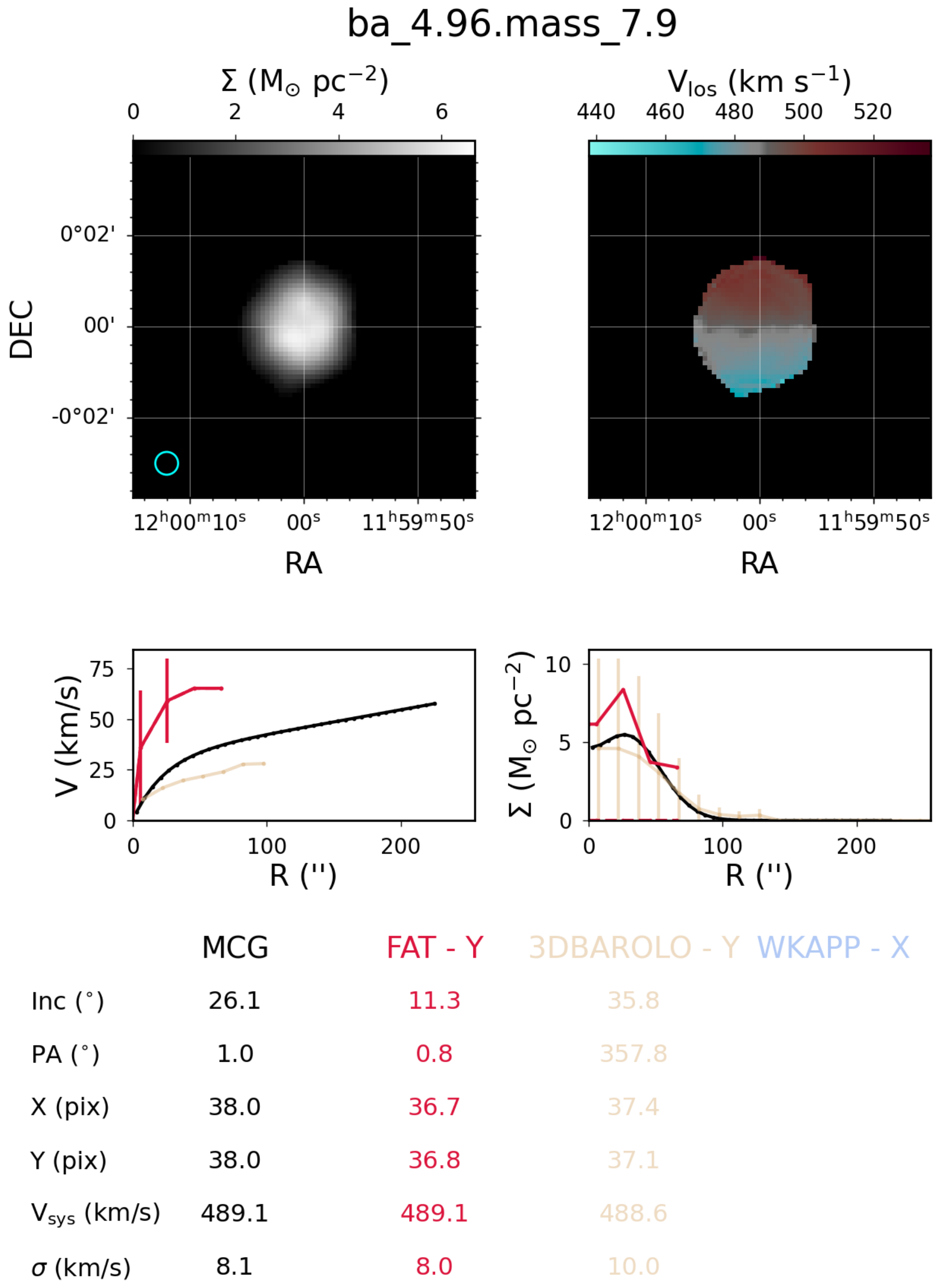} 
    \caption{An example diagnostic plot comparing \barolo, \fat, and \wkapp\ models to the underlying \mcgsuite\ model for a test mock galaxy used in Sec. \ref{Sec:LowVelModellingTests}.  The name of the test galaxy is listed at the top, which is named by the size (ba==beams across) and logarithmic mass.  The black, red, yellow, and blue (missing in this case) lines in the rotation curve and surface density panels (lower left and right panels respectively) show the underlying \mcgsuite\ model and the \fat\, \barolo\, and \wkapp\ models respectively.  Similarly, the values for geometric parameters (inclination, position angle, centroid, and systematic velocity) and the velocity dispersion for the underlying model and each fit are listed in the table below.  In this particular case, the various fits do not agree with the underlying model.}
  \label{Fig:FailFit}
\end{figure*}

\begin{figure*}
\centering
    \includegraphics[width=0.9\textwidth]{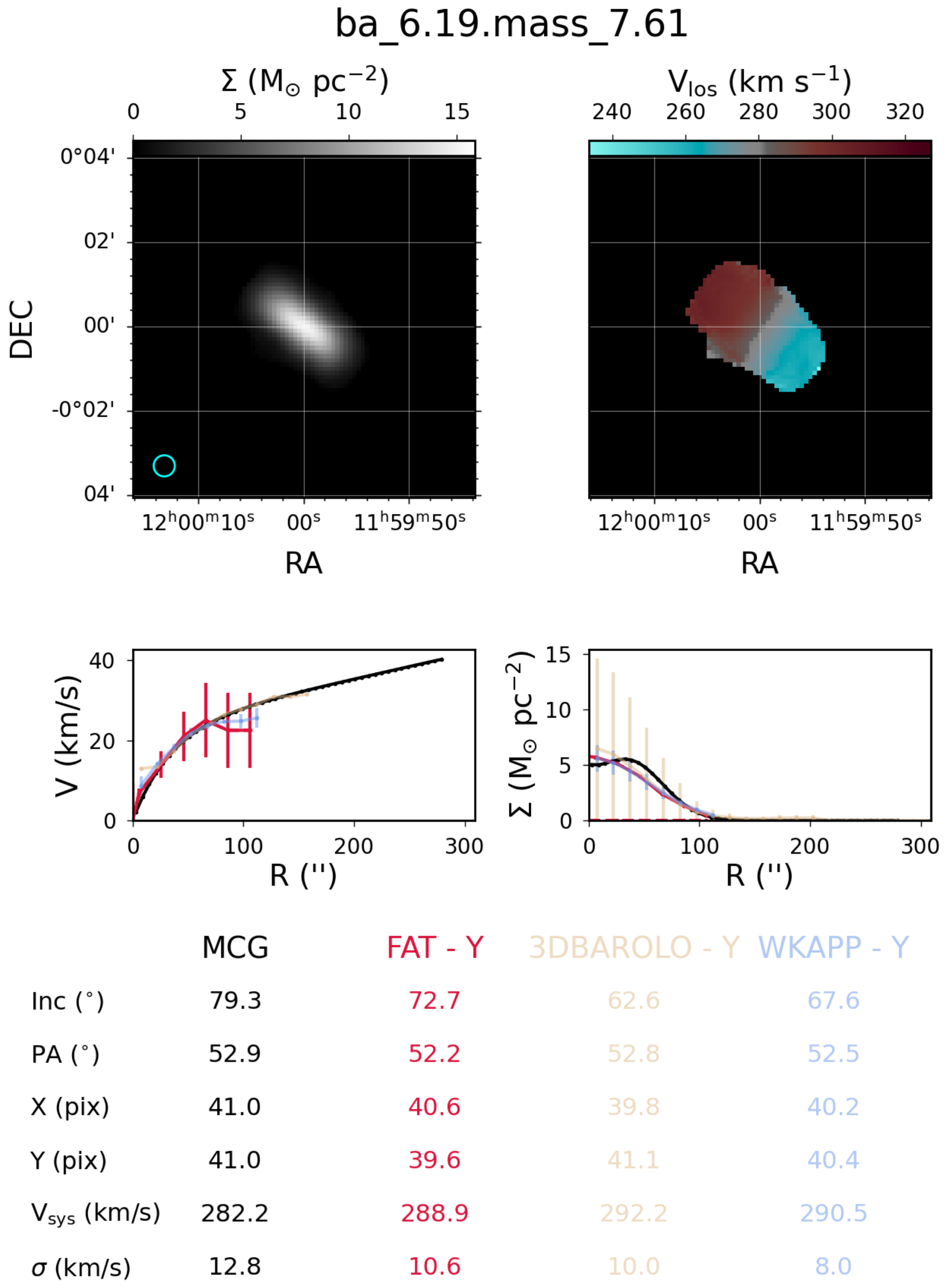} 
    \caption{An example diagnostic plot comparing \barolo, \fat, and \wkapp\ models to the underlying \mcgsuite\ model for a test mock galaxy used in Sec. \ref{Sec:LowVelModellingTests}.  The name of the test galaxy is listed at the top, which is named by the size (ba==beams across) and logarithmic mass.  The black, red, yellow, and blue lines in the rotation curve and surface density panels (lower left and right panels respectively) show the underlying \mcgsuite\ model and the \fat\, \barolo\, and \wkapp\ models respectively.  Similarly, the values for geometric parameters (inclination, position angle, centroid, and systematic velocity) and the velocity dispersion for the underlying model and each fit are listed in the table below.  In this particular case, all fits agree with the underlying model.}
  \label{Fig:SuccessFit}
\end{figure*}

In order to test the various fits in Sec. \ref{Sec:LowVelModellingTests}, we make diagnostic plots comparing the underlying \mcgsuite\ to the \barolo\, \fat\, and \wkapp\ for each test galaxy.  Figure \ref{Fig:FailFit} shows an example of these diagnostic plots where the \barolo\ and \fat\ fits are acceptable \textit{in isolation}, but, when compared to the underlying \mcgsuite\ model, they are visually determined to disagree with the truth.  Such models will appear as red circles in Figures \ref{Fig:LowVel_MCGResults}-\ref{Fig:UDG_MCGResults_HighRes}.  In this particular test case, the \fat\ and \barolo\ models differ enough from each other that \wkapp\ does not produce a model. By contrast,  Figure \ref{Fig:SuccessFit} shows a model where \fat\ and \barolo\ agree with other, a \wkapp\ model is produced, and all three agree with underlying \mcgsuite\ model.

\section{Obscured Galaxies}\label{app:Obs}

\begin{figure*}
\centering
    \includegraphics[width=0.9\textwidth]{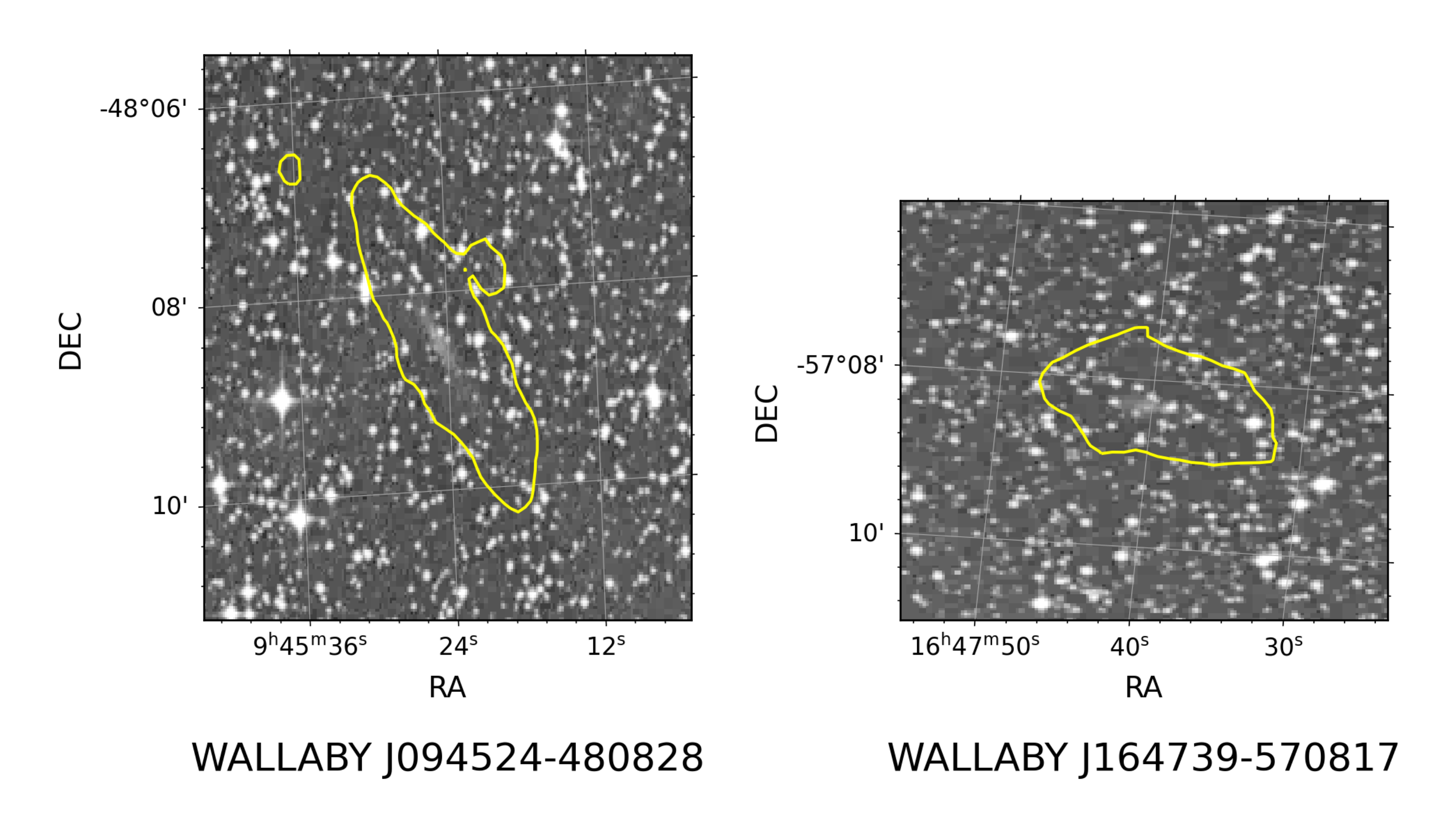} 
    \caption{The DSS-II blue images for the two galaxies without DECaLS data.  The yellow contour shows the $1~\Msol~\rm{pc}^{-2}$ surface density limit.}
  \label{Fig:Obs}
\end{figure*}

WALLABY J094524-480828 and WALLABY J164739-570817 both lie near the zone of avoidence and are not included in the DECaLS optical bands.  However, they have been observed as a part of DSS-II.  Figure \ref{Fig:Obs} shows the blue band images for these dwarfs overlaid with contours at $1~\Msol~\rm{pc}^{-2}$ from WALLABY.  We have used the blue bands as these show the detections most clearly.  Due to the data being both shallow and subject to large extinctions we have not generated three color images for these galaxies.   

\end{document}